\documentclass[apj,twocolumn]{aastex63}

\usepackage{graphicx}  
\usepackage{color}     
\usepackage{amsmath,amssymb}   
\usepackage{enumerate}
\usepackage{lipsum}
\usepackage{booktabs}
\usepackage{wrapfig}
\usepackage[makeroom]{cancel}
\usepackage[caption=false]{subfig}

\def\dr{\text{d}r}

\def\STIR/{\textit{STIR}}
\def\vturb{v_{\rm turb}}
\def\omegaBV{\omega_{\rm BV}}
\def\aMLT{\alpha_{\rm MLT}}
\def\BV/{Brunt-V\"{a}is\"{a}l\"{a}}

\newcommand{\pder}[2]{\frac{\partial#1}{\partial#2}}

\shorttitle{EOS and Turbulent CCSNe}
\shortauthors{Boccioli et al.}

\begin{document}
\doublespace

\title{Effect of the Nuclear Equation of State on Relativistic-Turbulence Induced Core-Collapse Supernovae}

\correspondingauthor{Luca Boccioli}
\email{lbocciol@nd.edu}

\author[0000-0002-4819-310X]{Luca Boccioli}
\affiliation{Center for Astrophysics, Department of Physics, University of Notre Dame, 225 Nieuwland Science Hall, Notre Dame, IN 46556, USA}

\author[0000-0002-3164-9131]{Grant J. Mathews}
\affiliation{Center for Astrophysics, Department of Physics, University of Notre Dame, 225 Nieuwland Science Hall, Notre Dame, IN 46556, USA}

\author[0000-0002-6923-6455]{In-Saeng Suh}
\affiliation{Center for Astrophysics, Department of Physics, University of Notre Dame, 225 Nieuwland Science Hall, Notre Dame, IN 46556, USA}
\affiliation{Center for Research Computing,  University of Notre Dame,   Notre Dame, IN 46556, USA}

\author[0000-0002-8228-796X]{Evan P. O'Connor}
\affiliation{The Oskar Klein Centre, Department of Astronomy, Stockholm University, AlbaNova, SE-106 91 Stockholm, Sweden}

\begin{abstract}
The nuclear equation of state is an important component in the evolution of core collapse supernovae.  In this paper we make a survey of various equations of state in the literature and analyze their effect on spherical core-collapse models in which the effects of three-dimensional turbulence is modeled by a general relativistic formulation of Supernova Turbulence in Reduced dimensionality (STIR).
We show that the viability of the explosion is quite EOS dependent and that it best correlates with the early-time interior entropy density of the proto-neutron star. We check that this result is not progenitor dependent, although low-mass progenitors show different explosion properties, due to the different pre-collapse nuclear composition. Larger central entropies also induce more vigorous proto-neutron-star convection in our one-dimensional turbulence model, as well as a wider convective layer.
\end{abstract}

\section{Introduction} 
\label{sec:intro}
The detailed explosion mechanism for core-collapse supernovae (CCSNe) has been one of the key problems in nuclear astrophysics for more than 50 years, and resolving this problem  still poses huge physical and computational challenges. The thermodynamic conditions typical of the collapse and subsequent explosion of massive stars are characterized by wide ranges of densities ($10^3-10^{15}$ g cm$^{-3}$), temperatures ($1-100$ MeV) and electron fractions ($0.01-0.6$). Therefore, a detailed knowledge of the properties of matter under all these thermodynamic conditions is required. However, the Equation of State (EOS) for nuclear matter, especially at high densities, is still poorly constrained (e.g. \cite{Klahn2006_HIC_constraints_EOS,Hebeler2013_constraints_EOS,Zhang2018_constraints_on_EOS,Burgio2021_constraints_on_EOS}). In addition to this, an extremely large number of neutrinos is produced during the collapse, some of which can be trapped inside the newly formed Proto-Neutron Star (PNS). Therefore, an accurate description of the interactions between neutrinos and regular matter is needed, as well as advanced algorithms describing how they are transported from the center to the outer layers of the star \citep{Liebendorfer2005_comparison_NT_methods,Kato2020_NT_MC_method_nucleon_recoils,Mezzacappa2021_NT_review}.

In this article, we focus on the impact that the EOS can have on the explosion mechanism of CCSNe by using the spherically symmetric, general relativistic model described in \cite{Boccioli2021_STIR_GR}. In that paper, the Supernova Turbulence In Reduced-dimensionality model (STIR), developed by \cite{Couch2020_STIR}, was extended to incorporate a general relativistic treatment of  turbulent convection.  Then, it was studied how turbulent convection behaves with and without taking General Relativity (GR) into account. Lastly, the effect that this has on the explodability as a function of progenitor mass was also analyzed. However, in that work only a single EOS was considered, and the properties of the proto-neutron star (PNS) were not discussed. In the present study, we extend that analysis by investigating the effects of the EOS on the explosion and on the properties of the PNS.

There have been a number of studies in the past regarding the impact of the nuclear equation of state on CCSNe (e.g. \cite{Lattimer1991_LS,OConnor2011_explodability,Hempel2012_SN_simulations,Janka2012_review_CCSNe,Steiner2013_SFHo,Couch2013_EOS_dependance,Suwa2013_EOS_dependance,Fischer2014_Sym_Ene_in_SN,Char2015_HyperonEOS_in_SN,Olson2016_NDL_EOS,Furuwawa2017_VM_vs_FYSS_EOSs,Richers2017_EOS_effect_on_GW_SN,Nagakura2018_2D_full_transpo_2EOS,Morozova2018_GW_SN_EOS,Burrows2018_Physical_dependencies_SN,Schneider2019,Harada2020_2D_SN_EOS_effect,Yasin2020_EOS_effects_1DFLASH}), but in most of those studies only one progenitor and a few EOSs were considered (although see the recent paper by \cite{Ghosh2021_PUSH_EOS}, where they study the impact of the EOS on the nucleosynthesis of several progenitors). Here, we study four progenitors modeled with a broad range of EOSs, some of which were constructed using Skyrme density functionals and others using Relativistic Mean Field (RMF) theory.

There exist other recent studies that have analyzed the impact of the EOS on the explosion properties and on the PNS structure \citep{Schneider2019,Yasin2020_EOS_effects_1DFLASH}.  However, they only focused on EOSs generated using Skyrme energy-density functionals, since they were interested in identifying how different nuclear parameters could impact the physics of CCSNe. In those studies, they found that the effective nucleon mass $m_n^*$ is correlated with the strength of the explosion. In this paper we take a different approach, and compare EOSs calculated using different theoretical frameworks, i.e. both Skyrme density functionals and relativistic mean field theory. In this way we are able to analyze how differences in the underlying theoretical framework translate into differences in explosion properties and PNS structure.

This is the first extensive EOS study based upon a relativistic explosion model driven by turbulent convection, and therefore we are also able to analyze the impact that the EOS has on the convection in the gain region and inside the PNS. In Section \ref{sec:methods} we describe the EOSs adopted, the numerical setup chosen and the essential physics of STIR. In Section \ref{sec:calibration} we describe how we calibrate our parametric model by comparing to 3D simulations. Results regarding the explosion properties, the structure of the PNS and neutrinos-related quantities are reported in Section \ref{sec:results}.  Conclusions are given in Section \ref{sec:conclusions}. Throughout the manuscript, we adopt natural units, i.e. $G = c = {\rm M}_\odot = 1$.

\section{Methods}
\label{sec:methods}
\subsection{Equations of State}
\label{sec:EOS-types}
To study the effect of the EOS on the explosion properties, we simulated the explosion of four representative progenitors with initial main-sequence masses of 9, 15, 20 and 25 M$_\odot$ from \cite{Sukhbold2016_explodability} using 7 different EOSs summarized in Table \ref{tab:EOS_properties}. We employ some of the most common EOSs used for CCSNe simulations to be consistent with previous results presented in the literature (e.g. \cite{Lattimer1991_LS,OConnor2011_explodability,Hempel2012_SN_simulations,Janka2012_review_CCSNe,Steiner2013_SFHo,Couch2013_EOS_dependance,Suwa2013_EOS_dependance,Fischer2014_Sym_Ene_in_SN,Char2015_HyperonEOS_in_SN,Olson2016_NDL_EOS,Furuwawa2017_VM_vs_FYSS_EOSs,Richers2017_EOS_effect_on_GW_SN,Nagakura2018_2D_full_transpo_2EOS,Morozova2018_GW_SN_EOS,Burrows2018_Physical_dependencies_SN,Schneider2019,Harada2020_2D_SN_EOS_effect,Yasin2020_EOS_effects_1DFLASH}), despite the fact that some of these equations of state are inconsistent with recent observational and/or experimental constraints.

The nuclear properties that can most significantly impact the dynamics of CCSNe and determine the properties of the proto-neutron star formed in the collapse \citep{Schneider2019,Yasin2020_EOS_effects_1DFLASH} are: (i) the zero temperature saturation density of symmetric nuclear matter $n_0$; (ii) the effective nucleon mass in symmetric matter at saturation density $m^*_n$; (iii) the nuclear incompressibility $K_0$; (iv) the density-dependent symmetry energy $\epsilon_{\rm sym}(n)$. The latter is defined as the iso-vector component of the energy per baryon, obtained by expanding the energy per baryon around its value for symmetric matter:

\begin{equation}
    \epsilon_B(n,y) =  \epsilon_B(n,y=1/2) + \epsilon_{\rm sym} \delta(y)^2 ~,
\end{equation}
where $y$ is the proton fraction and $\delta(y) = 1-2y$. Hence, the symmetry energy is defined as:

\begin{equation}
    \epsilon_{\rm sym} = \epsilon_B(n,y=1/2) - \epsilon_B(n,y=0).
\end{equation}
It is also common to encounter an alternative definition of the symmetry energy, which can be derived by expanding $\epsilon_B(n,y)$ around its value for symmetric matter (i.e. $\epsilon_B(n,y=1/2)$), yielding:

\begin{equation}
    \epsilon_{\rm sym} = \epsilon_B(n,y=1/2) + \epsilon_{\rm sym} \delta(y)^2 + \mathcal{O}[\delta(y)^4].
\end{equation}
The two definitions agree up to quartic terms. 

\begin{figure}
    \centering
    \includegraphics[width=\columnwidth]{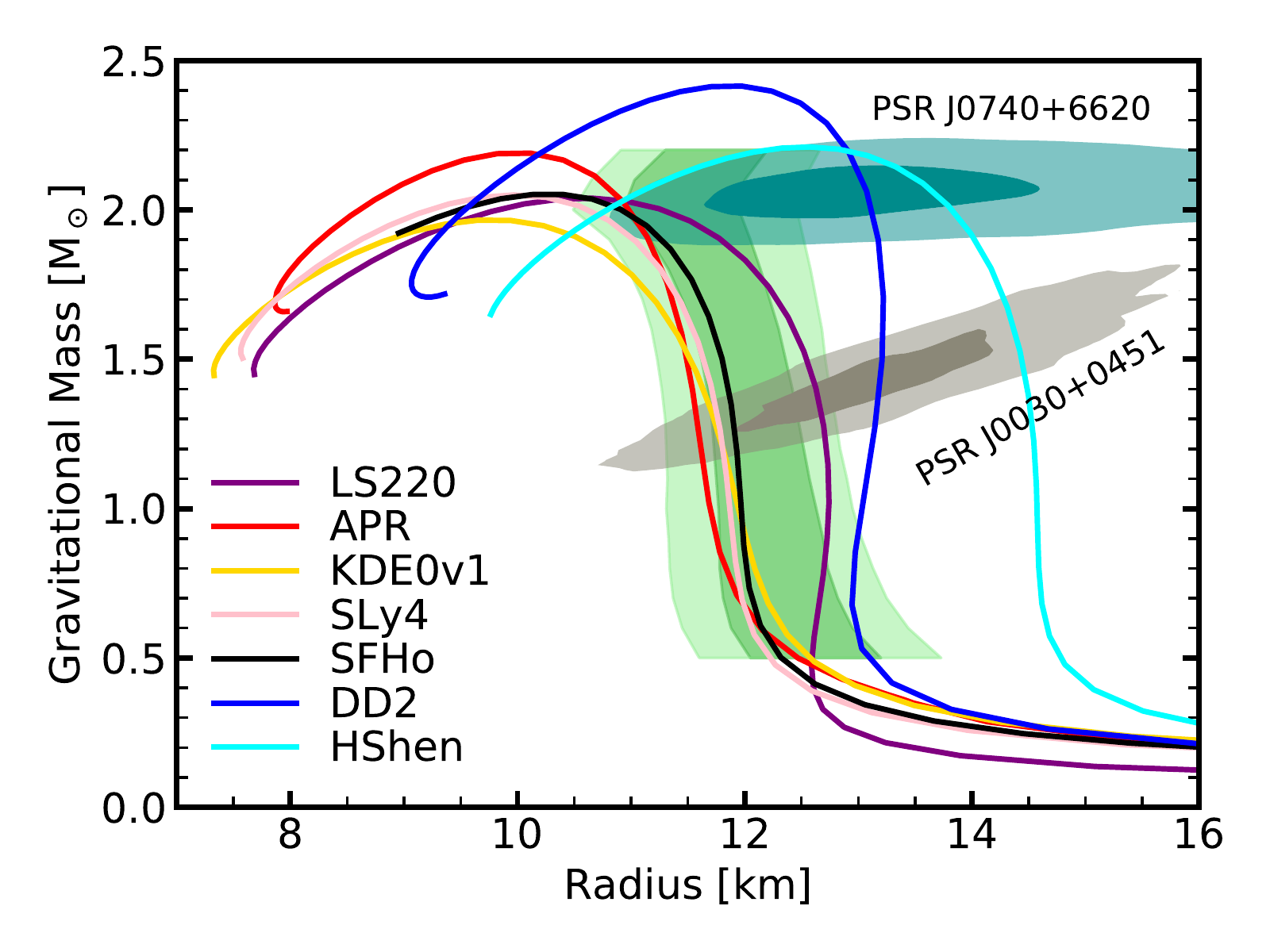}
    \caption{Gravitational mass vs radius, from the various equations of state adopted in the present work. The green shaded region shows the NS mass-radius constraints from model A of \cite{Nattila2016_EOS_constraints}. The grey contour represents the mass and radius of the millisecond pulsar \texttt{PSR J0030+0451} from \cite{Miller2019_NS_contours}, while the blue contour represents the mass and radius of the millisecond pulsar \texttt{PSR J0740+6620} from \cite{Miller2021_NS_contours}, both measured by the NICER collaboration.}
    \label{fig:mass_vs_radius}
\end{figure}

A brief summary of these equations of state is as follows: 
\begin{itemize}
    \item  The APR EOS  \citep{APR_1998,Schneider2019_APR_EOS} is based upon a thorough variational calculation, including realistic two- and three-body nuclear forces {\color{red} to derive the ground state energies of pure neutron matter and symmetric nuclear matter}. It utilizes a Hamiltonian that includes the two-body Argonne V18 potential extracted by fitting two-nucleon experimental scattering data, as well as the more empirical three-body Urbana IX potential, which is fit in part to light nuclei. This is the only EOS considered here that includes a phase transition to a neutral pion condensate. It is a generalization of the original EOS at zero temperature from \cite{APR_1998}, where the nuclear potentials are extended to the finite temperature part. At low densities it is connected to a network of $\sim$3300 nuclei in Nuclear Statistical Equilibrium (NSE) using the SRO code \citep{Schneider2017_SROEOS}.
    
    \item The LS220 \citep{Lattimer1991_LS}, KDE0v1 \citep{Agrawal2005_KDE0v1_ref_SRO} and SLy4 \citep{Chabanat1997_SLy4} parameterization are taken from \cite{Schneider2017_SROEOS}. For these EoSs  nuclear matter is described using Skyrme models for nuclear interactions, and in the transition to nuclear saturation density nuclei are described as a compressible liquid drop using the SNA. The version of LS220 utilized here is in excellent agreement with the original work by \cite{Lattimer1991_LS}. At low densities these equations of state are coupled to an NSE network of $\sim$3300 nuclei using the SRO code \citep{Schneider2017_SROEOS}.
    
    \item The DD2 and SFHo EOSs are calculated using the Relativistic Mean-Field (RMF) approach of \cite{Hempel2010_HS_RMF} with the parametrization of the interactions between nucleons from \cite{Typel2010_DD2} and \cite{Steiner2013_SFHo}, respectively. Nuclear matter is treated using an NSE approach for the transition from non-uniform to uniform nuclear matter. These two EOSs differ from the EOSs described above in that they do not  employ the SNA in the vicinity of the saturation density. Rather, several thousand nuclei are taken into account, whose masses and binding energies, when available, are taken from experiment.
        
    \item The HShen EOS from \cite{Shen2011_HShen} is also based on  RMF theory.  This EOS differs from the SFHo and DD2 in that it uses a Single Nucleus Approximation (SNA) based upon a Thomas-Fermi model to describe matter in the approach to nuclear matter density.
\end{itemize}
The first four EOSs were generated using the SROEOS code from \cite{Schneider2017_SROEOS}. Therefore, they share the same treatment of inhomogeneous phases at sub-nuclear densities.  This consists of a finite temperature compressible liquid-drop model in which heavy nuclei are described using a single-nucleus approximation.
All of the  adopted equations of state in Table \ref{tab:EOS_properties} can thus be grouped in two categories: SRO-type (the first four), and the RMF-type (the last three). The gravitational mass versus radius relationships for cold neutron are shown in Figure \ref{fig:mass_vs_radius}, alongside some observational constraints on the mass and radius of Neutron Stars. As can be seen, not all these EOSs are consistent with the current observational constraints. Nevertheless, these formulations should represent a reasonable gamut of possible equations of state. 
 
\begin{table*}
\centering 
\begin{tabular}{l|cccccccc}
\toprule
\toprule
EOS & Type &       $n_0$  &     $J$ &       $L$ &       $K_0$  &  $m^*_n/m_{\rm n}$ &  $m^*_p/m_{\rm p}$ & M$_{\rm max}$ \\
 & & (fm$^{-3}$) & (MeV baryon$^{-1}$) & (MeV baryon$^{-1}$) & (MeV baryon$^{-1}$) & - & - & M$_\odot$ \\
\midrule
\midrule
  LS220 &  SRO & 0.1549 &  28.61 &   73.81 &  219.85 &    1.0000 &    1.0000 & 2.04\\
    APR &  SRO &0.1600 &  32.59 &   58.47 &  266.00 &    0.6980 &    0.6987 & 2.19 \\
 KDE0v1 &  SRO &0.1646 &  34.58 &   54.70 &  227.53 &    0.7440 &    0.7443 & 1.97 \\
   SLy4 &  SRO &0.1595 &  32.00 &   45.96 &  229.90 &    0.6950 &    0.6953 & 2.05 \\
\midrule
   SFHo &  RMF &0.1583 &  31.57 &   47.10 &  245.40 &    0.7609 &    0.7606 & 2.05 \\
    DD2 &  RMF &0.1491 &  32.73 &   57.94 &  242.70 &    0.5628 &    0.5622 & 2.14 \\
  HShen &  RMF &0.1455 &  36.95 &  110.99 &  281.00 &    0.6340 &       - & 2.21 \\
\bottomrule
\bottomrule
\end{tabular}
\caption{Parameters characterizing the equations of state utilized in this study: $n_0$ is the nuclear saturation density, $J$ is the symmetry energy at saturation density, $L$ is the slope of the symmetry energy, $K_0$ is the nuclear incompressibility, $m_n^*$ is the nucleon effective mass and $m_n$ is the nucleon mass. The first four equations of state are calculated using the SRO code from \cite{Schneider2017_SROEOS}, while the last three are calculated using RMF theory. }
\label{tab:EOS_properties}
\end{table*}

\subsection{Numerical model}
\label{sec:SN_model}
All of the simulations presented in this paper were run using a modified version of the open-source code \texttt{GR1D} \citep{OConnor2010,OConnor2015}. The original code is based on general relativistic, spherically symmetric hydrodynamics and radiation transport. The Boltzmann equation that describes the propagation of neutrinos is solved with a two-moment scheme, the so-called M1 transport \citep{Shibata2011,Cardall2013_GR_M1}, while neutrino opacities were generated using the open-source code \texttt{NuLib} \citep{OConnor2015}. 

We use the same spatial grid for all the simulations, i.e. 700 zones with a constant resolution of 0.3 km in the inner 20 km and then logarithmically spaced outer zones extending to a radius of $15\,000$ km. The only exception are the progenitors with very steep density gradients (i.e. the ones with masses smaller than 12 M$_\odot$), for which we place the outer boundary at the radius where the density reaches $6 \times 10^4\ {\rm g\ cm}^{-3}$, which can vary between $5\,000$ km and $10\,000$ km. 

For the neutrinos, we use 18 energy groups logarithmically spaced from 1 to 280 MeV. We adopted the opacities described in \cite{Bruenn1985}, with weak-magnetism and recoil corrections from \cite{Horowitz2002} and a virial correction to neutrino-nucleon scattering from \cite{Horowitz2017_virial}. We include inelastic neutrino-electron scattering and velocity-dependent terms in the transport equation according to \cite{OConnor2015}.

\subsection{Spherical relativistic turbulence model}
Neutrino heated turbulence plays a crucial role in the shock revival and explodability, as many previous studies have noted \citep{Couch2015_turbulence, Radice2016, Radice2018_turbulence, Mabanta2018_MLT_turb}. In essence, material behind the expanding shock is heated by neutrinos escaping the proto-neutron star.  This increases the entropy and leads to a negative entropy gradient which triggers turbulent convection. The rising turbulent fluid elements can then transfer energy to the shock.  This makes spherical  explosions triggered by neutrino heated turbulent convection physically well-motivated.

To analyze the impact of each EOS on the explosion, we used a modified version of \texttt{GR1D} described in \cite{Boccioli2021_STIR_GR} based upon the Supernova Turbulence In Reduced-dimensionality (STIR) model of \cite{Couch2020_STIR}. With that, one can include the effects of turbulent convection in our spherically symmetric simulations by using a time-dependent Mixing-Length Theory (MLT) approach, and therefore trigger an explosion. 

The difference between our model and the original STIR is that we explicitly take General Relativity (GR) into account. Despite yielding small differences in the shock dynamics, GR  can have a significant effect in the explodability of SNe as a function of progenitor mass \citep{Boccioli2021_STIR_GR}. Here we summarize the main features of this model, but more details can be found in \cite{Mabanta2018_MLT_turb}, \cite{Couch2020_STIR} and \cite{Boccioli2021_STIR_GR}.
The metric adopted in \texttt{GR1D} is the following:

\begin{equation}
    \label{eq:metric}
    \begin{split}
    ds^2 &= g_{\mu\nu} x^\mu x^\nu \\
         &= -\alpha(r,t)^2 dt^2 + X(r,t)^2 dr^2 + r^2 d\Omega^2 ~,
    \end{split}
\end{equation}
where the $t-t$ component $\alpha$ is the lapse function and the $r-r$ component $X$ is a function of the gravitational mass of the system at radius $r$.

In addition to  the standard hydrodynamic equations solved in \texttt{GR1D} \citep{OConnor2010}, we have added an equation to account for the time evolution of the turbulent energy:

\begin{equation}
\begin{split}
\label{eq:vturb}
&\pder{D \vturb^2}{ t} + \frac{1}{r^2}\pder{}{r} [\frac{\alpha r^2}{X} D(\vturb^2 v - D_{\rm K} \nabla \vturb^2)] \\ 
&= -\alpha X \left(\rho\vturb^2 \pder{v}{r} + \rho\vturb \omegaBV^2 \Lambda_{\text{mix}} - \rho\frac{\vturb^3}{\Lambda_{\text{mix}}}\right) ~,
\end{split}
\end{equation}
where $\vturb$ is the turbulent velocity and $D = X \rho W$ is the conserved density, with $W = (1 - v^2)^{-1/2}$ being the Lorentz factor. The proper physical velocity is defined as $v = X v^r$, where $v_r$ is the coordinate velocity, and $D_{\rm K} = \alpha_{\rm K} \vturb \Lambda_{\text{mix}}$ is a diffusion coefficient. The two main quantities characterizing the model are the mixing length

\begin{equation}
    \label{eq:lambda_mix}
    \Lambda_{\text{mix}} = \aMLT \frac{P}{\rho ~\text{d}\phi/\dr} ~,
\end{equation}
and the \BV/ frequency

\begin{equation}
    \label{eq:BV_GR}
    \omegaBV^2 =  \frac{\alpha^2}{\rho h X^2} \left(\frac{\text{d}\phi}{\dr} - v \pder{v}{r}\right)\left(\pder{\rho(1+\epsilon)}{r} - \frac{ 1}{c_s^2} \pder{P}{r} \right) ~,
\end{equation}
where $\phi=\ln \alpha$ reduces to the Newtonian potential in the non-relativistic limit. The quantity  $h = 1 + \epsilon + P/\rho$ is the relativistic enthalpy, where $\epsilon$ is the internal energy, $P$ is the pressure, and $c_s$ is the sound speed. 

The above equations depend on two parameters: $\aMLT$ and $\alpha_{\rm K}$. Additional diffusion coefficients analogous to $D_{\rm K}$ are added to the evolution equations for internal energy, electron fraction and neutrino energy density, increasing the number of parameters in the model to 5. However, we fix $\alpha_{\rm K}$ and the other mixing length parameters  to $1/6$, consistent with the choice of \cite{Muller2016_Oxburning} and \cite{Couch2020_STIR}, since the model is not very sensitive to their value. The only parameter left is $\aMLT$, which determines the magnitude of the mixing length and therefore the scale of convection.

\section{Calibration of the turbulence model}
\label{sec:calibration}
The size of the convective cells in our model depends upon the mixing length $\Lambda_{\rm  MLT}$, which is proportional to the mixing length parameter $\alpha_{\rm MLT}$. The larger $\alpha_{\rm MLT}$, the larger the convective eddies and the associated turbulent energy, which can therefore lead to a stronger shock revival. Following the approach of \cite{Couch2020_STIR}, we calibrate the value of $\alpha_{\rm MLT}$ by comparing to realistic convection in 3D simulations. As pointed out in \cite{Boccioli2021_STIR_GR}, the calibrated value of $\alpha_{\rm MLT}$ can vary by up to a few percent depending upon  spatial resolution and the number of neutrino energy groups adopted. Therefore, we maintain the same spatial resolution and number of neutrino energy groups in all our simulations, as described in Section \ref{sec:SN_model}.

For the calibration of $\alpha_{\rm MLT}$ we utilize the 20 M$_\odot$ solar-metallicity progenitor of \cite{Farmer2016}, and we employ the SFHo equation of state.  We then compare our simulations using 6 different values of $\alpha_{\rm MLT}$ with the 3D model of \cite{Couch2014_3D}, which utilized the same progenitor, the same EOS and a very similar M1 scheme for the neutrino transport. 

\begin{figure}
    \centering
    \includegraphics[width=\columnwidth]{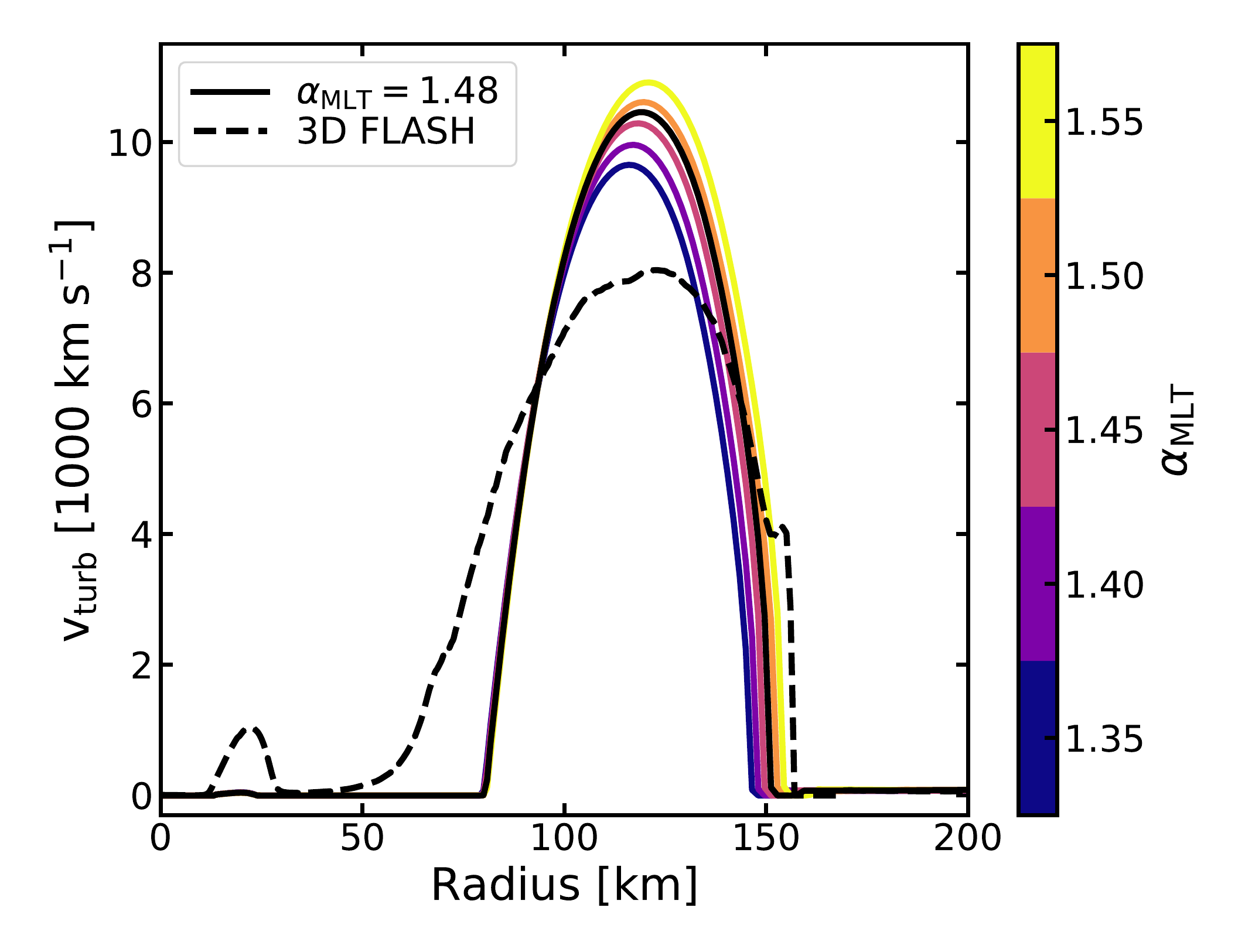}
    \caption{Turbulent velocity $\vturb$ as a function of radius for different values of $\aMLT$, taken at $\sim135$ ms after bounce. The dashed line represents the results from the 3D simulation of \cite{Couch2014_3D}, performed with FLASH.}
    \label{fig:calibration_vturb}
\end{figure}

A comparison of the turbulent velocities at $\sim$ 135 ms after bounce is shown in Figure \ref{fig:calibration_vturb}. As already pointed out in \cite{Couch2020_STIR}, the 3D profile exhibits a longer tail of convective motion between 50 and 75 km.  This results from non-radial motions. Indeed, the width of the gain region in 3D simulations varies as a function of angle.  As a consequence, the turbulent velocity can extend down to smaller radii. Lastly, the 3D profile shows a significant amount of convection in the PNS.  This convection is present but the convective velocity is much reduced in the 1D results.  We will come back to this in Section \ref{sec:PNS_interior}.

\begin{figure*}
    \centering
    \gridline{\fig{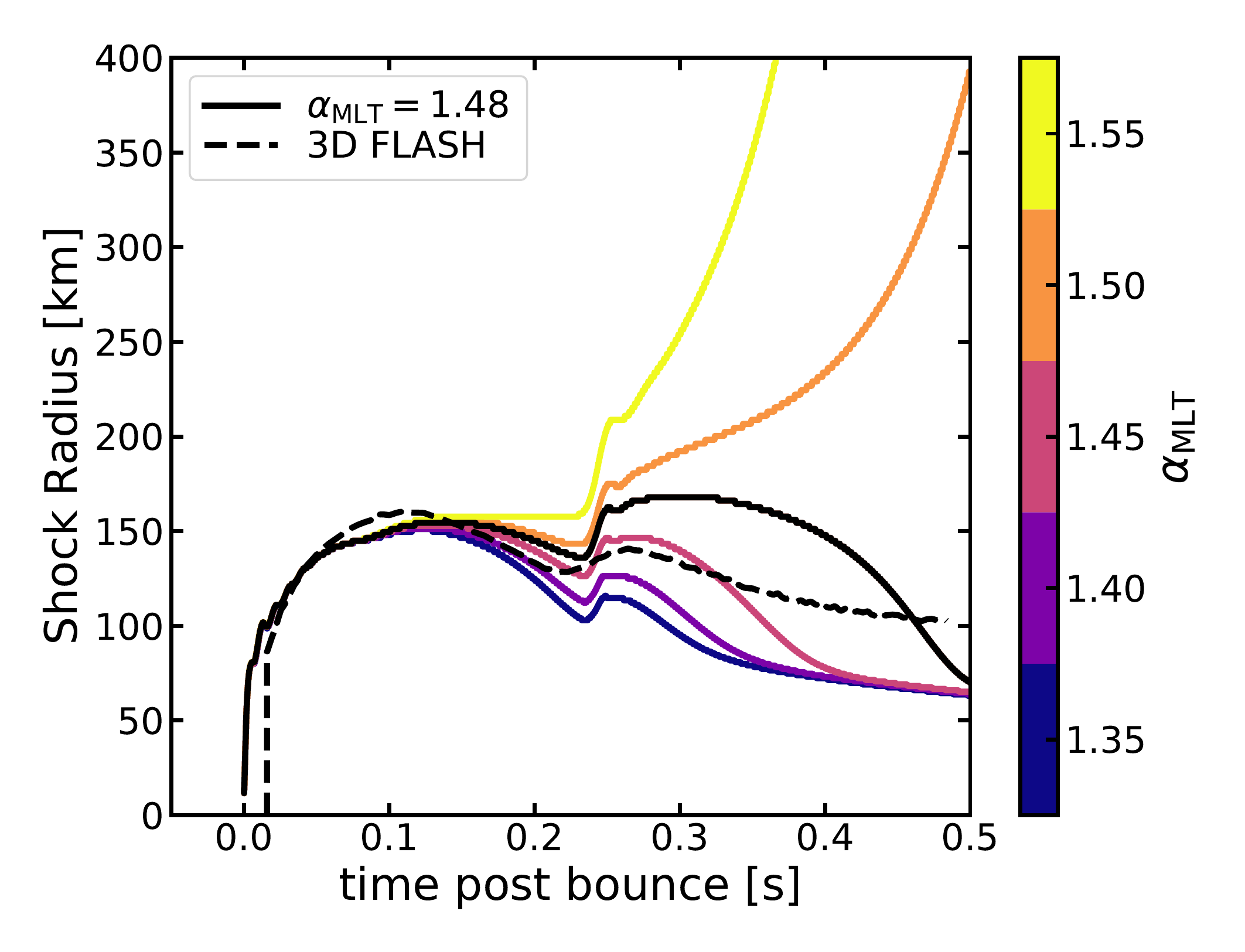}{0.5\textwidth}{(a)}
              \fig{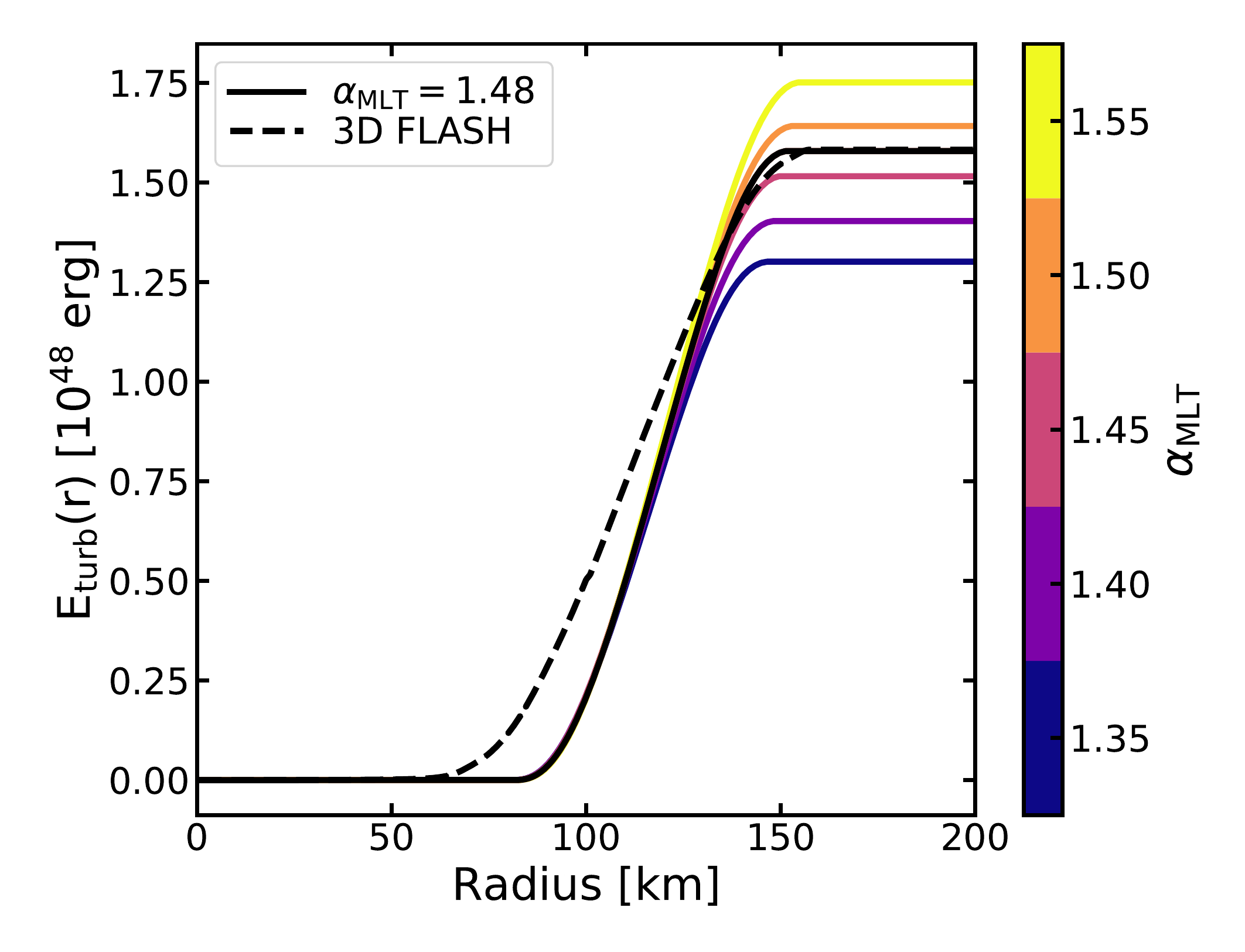}{0.5\textwidth}{(b)}}
    \caption{Panel (a) shows the shock radius as a function of time post bounce for different values of $\aMLT$. Panel (b) shows the integrated turbulent energy as a function of radius for a snapshot at $\sim$135 ms post bounce. Notice that, given its definition in Eq. \ref{eq:integrated_turbE}, $E_{\rm turb} (\rm r)$ is zero below 30 km and then it's constant for radii larger than $\sim 150$ km, which is approximately the location of the shock, above which no convection is developed.}
    \label{fig:calibration_shk_turbE}
\end{figure*}

To complete the calibration, we have analyzed two more quantities shown in Figure \ref{fig:calibration_shk_turbE}.  The first is the shock radius as a function of time and the second is the  integrated turbulent energy at $\sim$ 135 ms after bounce. 

Since convection in the PNS appears to be different in the 3D and 1D cases, we do not include it in our calibration analysis. Hence, we define the integrated turbulent energy as:

\begin{equation}
    \label{eq:integrated_turbE}
    E_{\rm turb}(r) = \int_{R^*}^r \rho v_{\rm turb}^2 r^{\prime 2} {\rm d}r^{\prime 2} ~,
\end{equation}
where we set $R^*=30$ km to exclude the convection in the PNS.

From the comparison shown in Figure \ref{fig:calibration_shk_turbE}, the value of $\alpha_{\rm MLT}$ that most closely matches the 3D results appears to be in the range between 1.45 and 1.5. Specifically, for $\aMLT = 1.48$, the integrated turbulent energy at radii $r \gtrsim 150$ km is in excellent agreement with the 3D case (see the right panel of Figure \ref{fig:calibration_shk_turbE}). In other words, at $\sim135$ ms after bounce, our simulation with $\aMLT = 1.48$ exhibits the same total turbulent energy in the gain region as the 3D case.

The excellent agreement with the total turbulent energy comes at the cost of a slightly larger shock radius in our 1D simulation compared to the 3D at times later than $\sim 250$ ms. The same discrepancy in the shock radius versus time can be found in the original \STIR/ simulations by \cite{Couch2020_STIR}. However, we choose to normalize the convection to the total turbulent energy at earlier times ($\sim 135$ ms) as this is when convection sets in and the outcome of the explosion is determined. One should also keep in mind that the shock radius in our 1D model tends to be a little larger than in the 3D case in the vicinity of the shock (see Figure \ref{fig:calibration_vturb}). This is because the turbulent velocity is only in the radial directions and therefore delivers energy to the outgoing shock more efficiently.  Nevertheless, we adopt this as the best representation of the contribution of neutrino-heated turbulence to the shock even though this can result in slightly larger shock radii for our 1D simulation.

\begin{figure}
    \centering
    \includegraphics[width=\columnwidth]{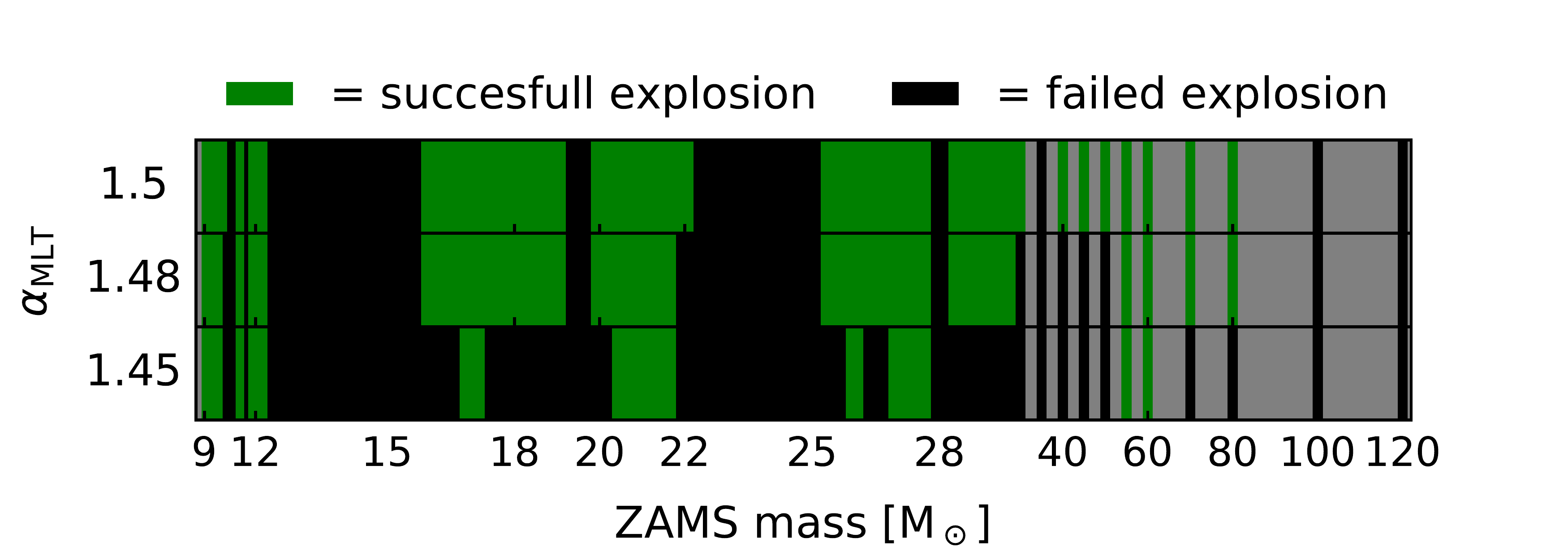} 
    \caption{Explodability of progenitors as a function of mass for different values of $\aMLT$. Green bands represent successful explosions (defined as those simulations for which the shock crosses a radius of 500 km), while black bands represent failed explosions. We color in grey the progenitors we did not simulate.}
    \label{fig:calibration_pattern}
\end{figure}

Given the discrepancy between 1D and 3D results seen in some of the hydrodynamical variables  analyzed above, we also analyze how the value of $\alpha_{\rm MLT}$ impacts the explodability as a function of progenitor mass. We have simulated 58 progenitors from \cite{Sukhbold2016_explodability}, with masses ranging from 9 to 120 M$_\odot$, using the SFHo EOS for three values of $\aMLT$: 1.45, 1.48, and 1.5. The outcome of these simulations is shown in Figure \ref{fig:calibration_pattern}. The explodability with $\alpha_{\rm MLT} = 1.48$ is consistent with the previous findings reported in \cite{Boccioli2021_STIR_GR}.

\begin{figure}
    \centering
    \includegraphics[width=\columnwidth]{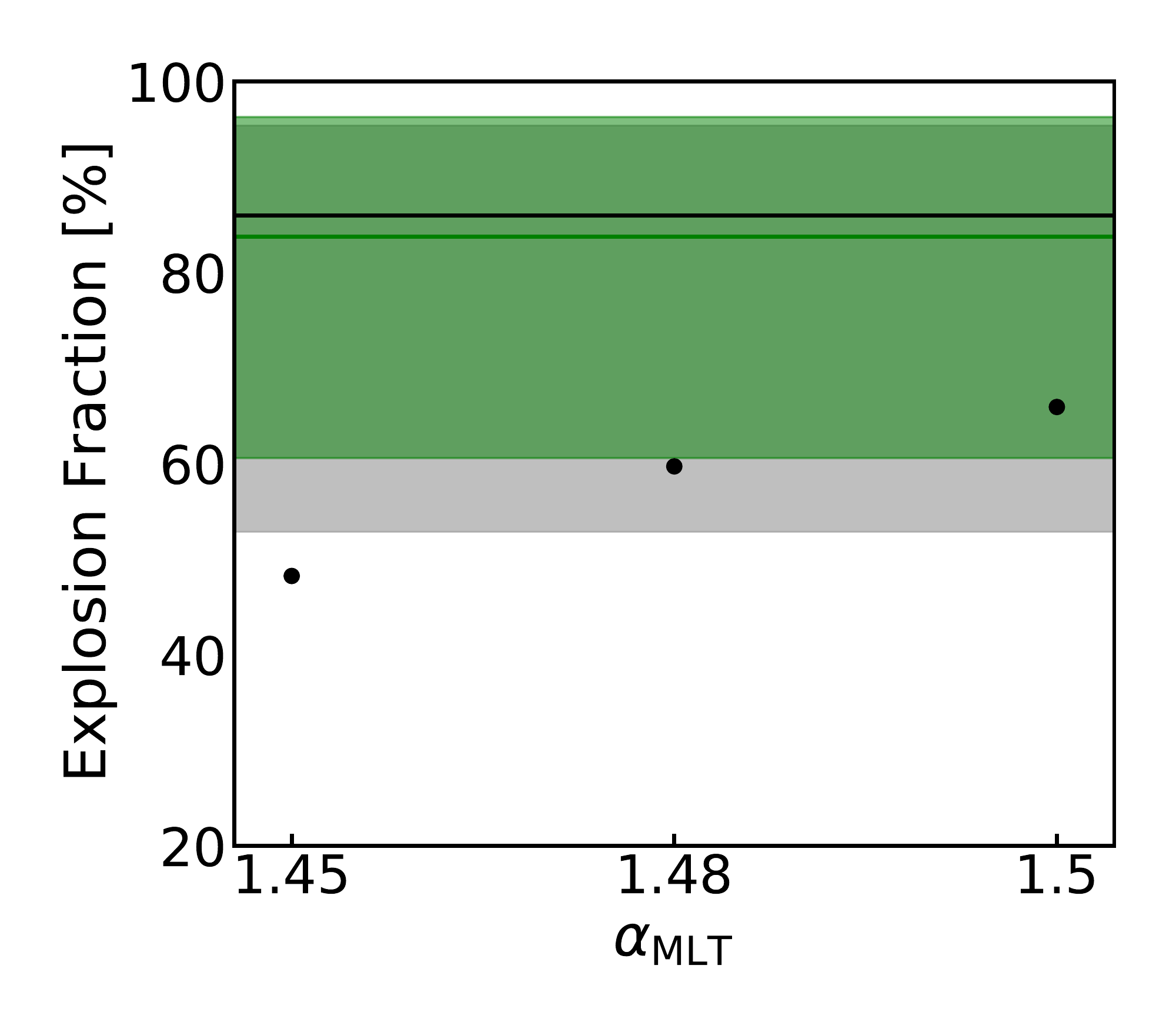} 
    \caption{The explosion fraction is calculated by weighing the explodability as a function of progenitor mass shown in Figure \ref{fig:calibration_pattern} with a Salpeter initial mass function. The black dots are the results from our simulations for different values of $\aMLT$. The black and green horizontal lines show the median from \cite{Adams2017_fraction_failed_SN} and \cite{Neustadt2021_failedSN_frac}, respectively. The shaded regions represent the 90\% confidence interval. Notice that in those papers they actually report the fraction of failed SN $f_{\rm FSN}$, therefore here we assume that the fraction of explosion is $1 - f_{\rm FSN}$ }
    \label{fig:calibration_expl_frac}
\end{figure}

With these results, one can calculate the total explosion fraction as a function of $\aMLT$, weighed by a Salpeter initial mass function. This is shown in Figure \ref{fig:calibration_expl_frac}. For comparison, we also show the observationally deduced range from \cite{Adams2017_fraction_failed_SN} (grey area) and the more recent results of \cite{Neustadt2021_failedSN_frac} (green area).

The explosion fraction for $\aMLT=1.48$ lies just outside the 90\% confidence interval of the most recent results, while for $\aMLT=1.5$ it is fully consistent with observations. However, since our hydrodynamical quantities are better reproduced by $\aMLT=1.48$ and the discrepancy with observational results is quite small, we adopt this value for the rest of the paper, although the same discussion and results are equally valid for simulations done with $\aMLT=1.5$.

After calibrating the value of $\aMLT$ we simulate four progenitors with masses of 9, 15, 20 and 25 M$_\odot$ for all of the EOSs listed in Section \ref{sec:EOS-types}.

\section{Results}
\label{sec:results}
\subsection{Explosion properties}
\label{sec:explodability}
\begin{figure*}
    \centering
    \includegraphics[width=\textwidth]{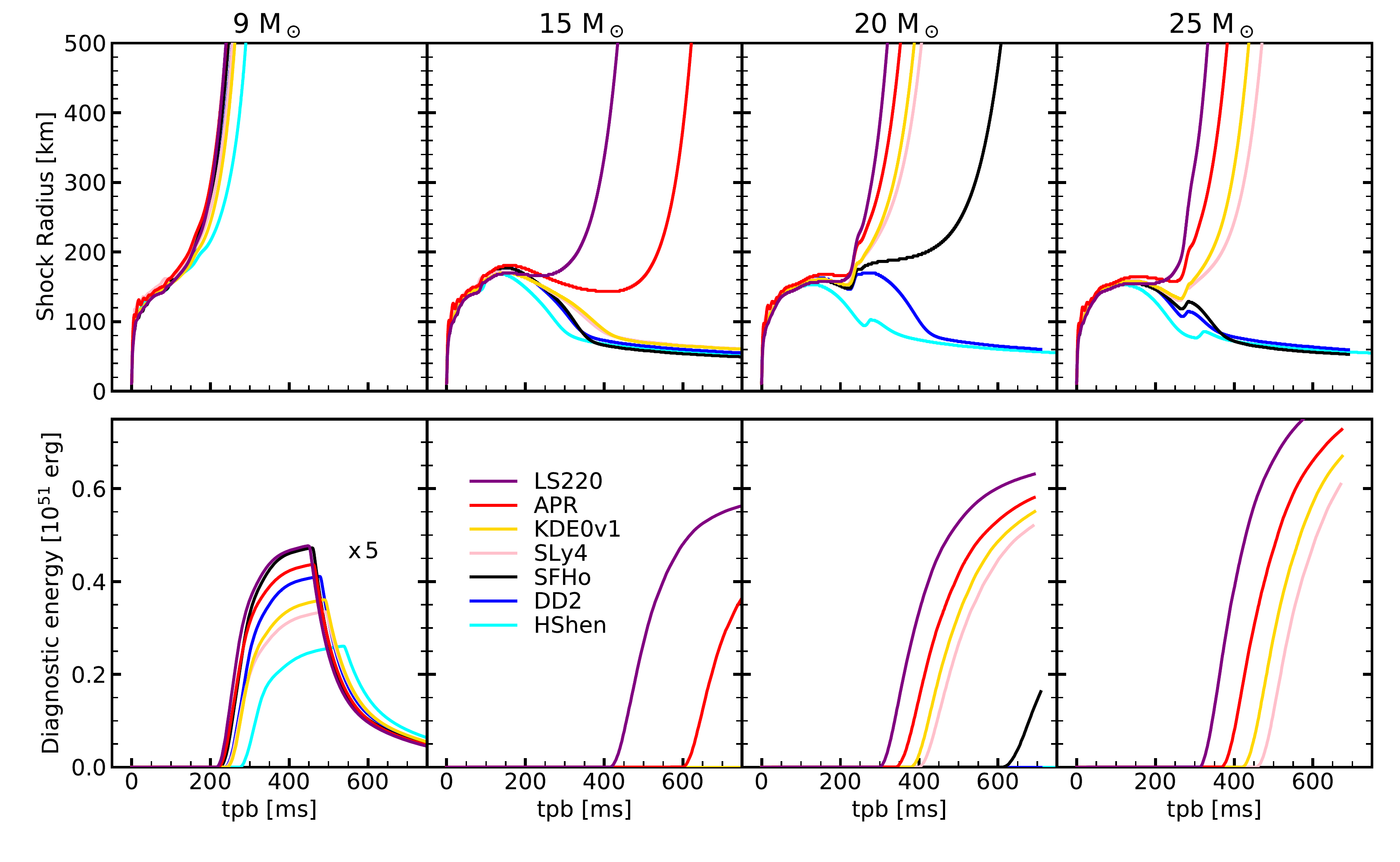}
    \caption{The upper panels display shock radius vs time for different progenitors and EOSs, while the lower panels display the diagnostic energy vs time. Notice that for the 9 M$_\odot$ progenitor the diagnostic energies have been multiplied by a factor of 5 so that they could be more clearly shown on the same scale used for the other progenitors.}
    \label{fig:shk_expl_ene}
\end{figure*}

Two of the main quantities of interest for an exploding supernova are the trajectory of the shock as a function of time and the explosion energy.  We display these in Figure \ref{fig:shk_expl_ene} for the four progenitors described in Section \ref{sec:intro}. Since the simulations we performed extend to less than one second of the post bounce phase, a  calculation of the final total explosion energy is not possible. Instead, one can define a diagnostic explosion energy \citep{Buras2006_2D_diag_ene,Marek2009_SASI_diag_ene,Muller2012_2D_GR,Bruenn2016_expl_en} as the integral of the binding energy $E_{\rm bind}$ in regions where $E_{\rm bind} > 0$. 

In the GR case, following \cite{Muller2012_2D_GR}, one can define $E_{\rm bind}$ in terms of the lapse function $\alpha$, the density $\rho$, the specific internal thermal energy $\epsilon_{\rm th}$, the pressure $P$ and the Lorentz factor $W$:

\begin{equation}
    E_{\rm bind} = \alpha\left[ ( \rho + \rho\epsilon_{\rm th} + P) W^2 - P\right] - \rho W ~.
\end{equation}

Here, it is important to distinguish between the traditional internal energy adopted in classical thermodynamics and the internal energy used in the context of CCSNe (see Appendix A of \cite{Bruenn2016_expl_en}). The latter definition incorporates the binding energy of nuclei in the expression of the internal energy.  This allows negative values of  $\epsilon$, hence making the definition of the total energy of the fluid problematic. A way around this is to define $\epsilon_{\rm th}$ as the difference between the internal energy $\epsilon$ --- taken from the EOS at a given density, temperature and electron fraction --- and the internal energy $\epsilon_0$, defined as the internal energy at zero temperature and at the same density and electron fraction\footnote{From a practical point of view, what we do is to take the internal energy corresponding to the lowest temperature of the table (typically $\sim 0.01$ MeV), which for the thermodynamic conditions of the supernova can be safely treated as zero.}. 

From this definition, it is clear that $\epsilon_0$ represents the binding energy of the nuclei, which at low temperature is the only significant contribution to the internal energy. Finally, the diagnostic explosion energy can be defined as:

\begin{equation}
    E_{\rm diag} = \int_{E_{\rm bind} >0} E_{\rm bind} {\rm d}V ~,
\end{equation}
where d$V$ is the proper volume.

As can be seen from the shock radius and the diagnostic energy, the LS220 EOS gives the earliest and strongest explosion, followed by the APR, KDE0v1, SLy4, SFHo, DD2 and HShen. This hierarchy is maintained across all progenitors, with the only exception being the 9 M$_\odot$ progenitor, for which the hierarchy of the explosion energies is different. Specifically,  the SFHo and the DD2 generate comparatively stronger explosions in the 9 M$_\odot$ progenitor than in the more massive progenitors, as can be seen from the diagnostic energy for this progenitor. 

In addition to this, it should be noted that the 9 M$_\odot$ progenitor leads to very early and faint explosions (the diagnostic energy in Figure \ref{fig:shk_expl_ene} has been multiplied by 8 so that it could more clearly be represented on the plot), and the decrease around 450 ms post bounce is a numerical artifact due to the shock leaving the outer boundary of the simulation. This trend for the 9 M$_\odot$ progenitor relates to the fact that these stars explode as  electron capture supernovae \citep{Isern1991_ECSNe}. A more detailed analysis of the differences between this progenitor and the more massive ones is given in Appendix \ref{sec:prog9}.  For now we will focus on the 15, 20 and 25 M$_\odot$ progenitors.

\begin{figure*}
    \centering
    \includegraphics[width=\textwidth]{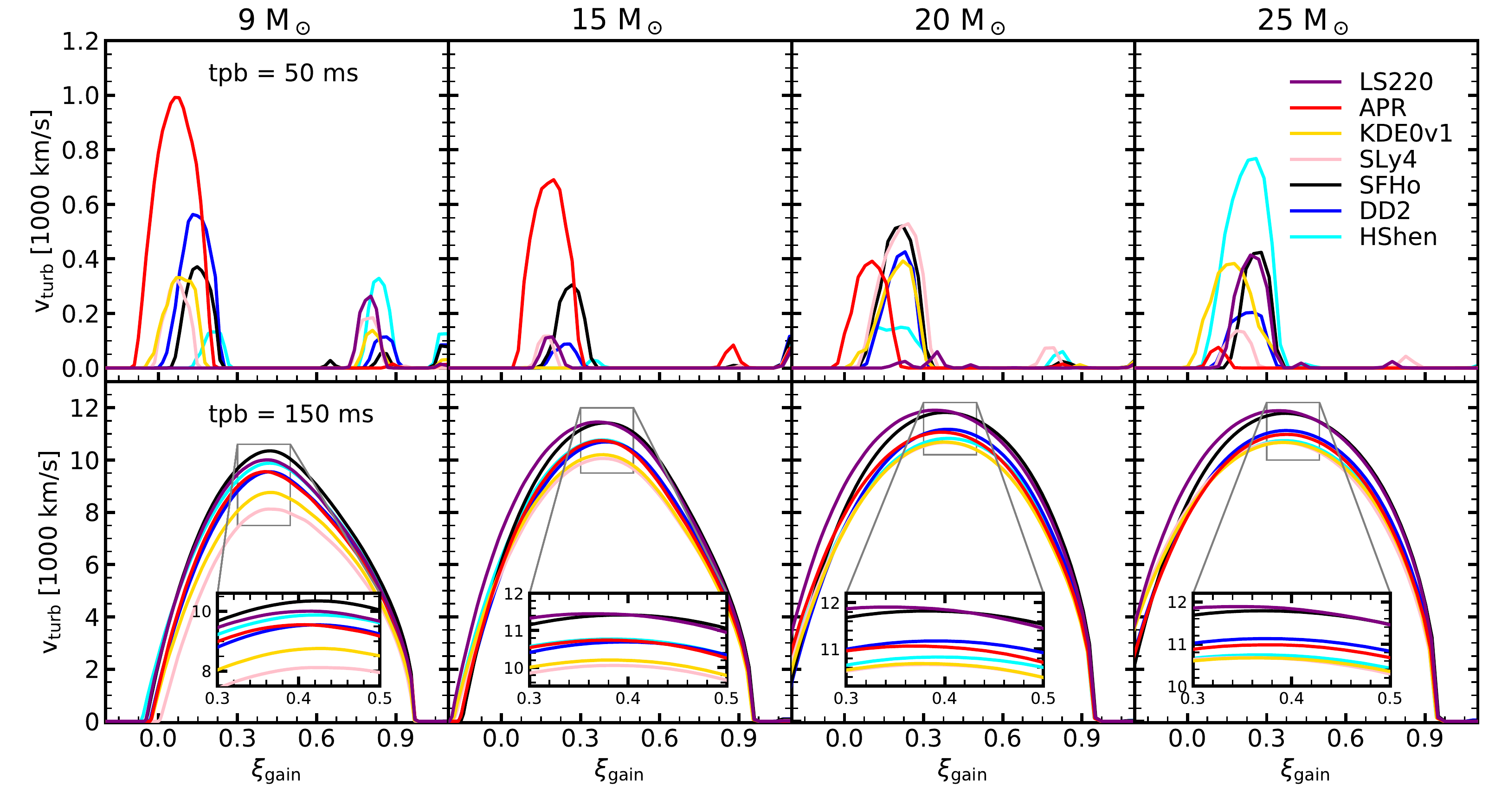}
    \caption{Turbulent velocity $\vturb$ in the gain region for different progenitors (columns) and EOSs (colors). The horizontal axis is the ``normalized radius" defined in Eq. \ref{eq:eps_gain} which is 0 at $r_{\rm gain}$ and 1 at $r_{\rm shk}$. The upper panels show profiles of turbulent velocity taken at $~50$ ms after bounce, when convection sths developing, while the lower panels are taken at $~150$ ms after bounce, when convection is fully developed.}
    \label{fig:turb_vel_profiles}
\end{figure*}

Since our convection model is designed to reproduce turbulence in the gain region, one can analyze how the explosion is affected by this mechanism. To make the comparison among different EOSs, which have different gain and shock radii, we define a dimensionless quantity:

\begin{equation}
    \xi_{\rm gain} = \frac{r - r_{\rm gain}}{r_{\rm shk} - r_{\rm gain}} ~,
    \label{eq:eps_gain}
\end{equation}
such that $\xi_{\rm gain} = 0$ when $r = r_{\rm gain}$ and $\xi_{\rm gain} = 1$ when $r = r_{\rm shock}$. The profiles of the turbulent velocity in the gain region are plotted in Figure \ref{fig:turb_vel_profiles} for two different time snapshots. 

At $\sim 50$ ms post bounce the gain region forms and the turbulent velocity starts building up. After that, as can be seen in the snapshot at $\sim 150$ ms, the profiles become smoother and the peak velocity steadily grows. Overall, the turbulent velocity profiles do not significantly change with progenitors and EOSs. 

Nevertheless, some global trends can be identified, since for example, the LS220 and the SFHo consistently yield the largest peak turbulent velocities, while the smallest corresponds to the SLy4. By looking at Figure \ref{fig:shk_expl_ene}, however, it is clear that the strongest explosions are given by the LS220 and the APR, which yields a smaller peak turbulent velocity than the SFHo. Hence, there doesn't seem to be any direct correlation between the strength of convection in the gain region and the strength of the explosion as a function of EOS.

It is interesting to note that if one only considers SRO-type EOSs or only RMF-type EOSs, the hierarchy of turbulent velocities follows the hierarchy of explosion strengths. Namely, for SRO-type EOSs, from largest to lowest peak turbulent velocity, we find: LS220, APR, KDE0v1, SLy4, which is the same as the hierarchy for explosion strengths. For RMF-type EOSs the hierarchy of peak turbulent velocities reads: SFHo, DD2, HShen, which is also the same as the hierarchy of explosion strength. 

Putting SRO and RMF types together, however, disrupts the correlation between explosion strength and convective velocity, as can be seen from the zoomed-in panels of Figure \ref{fig:turb_vel_profiles}. There, one can see that SFHo, DD2 and HShen yield a quite large turbulent velocity compared to the SRO-type EOSs, despite having low explosion energies, or not exploding at all. We will come back to this when we discuss neutrino luminosities in Section \ref{sec:neutrinos}. We suggest that this  difference in the turbulent velocity profiles (and neutrino luminosities, as discussed later) between SRO-type and RMF-type EOSs lies most likely in the different treatment of nuclear matter as it approaches saturation density.

\subsection{PNS interior and the role of central entropy}
\label{sec:PNS_interior}
\begin{figure*}
    \centering
    
    \includegraphics[width=\textwidth]{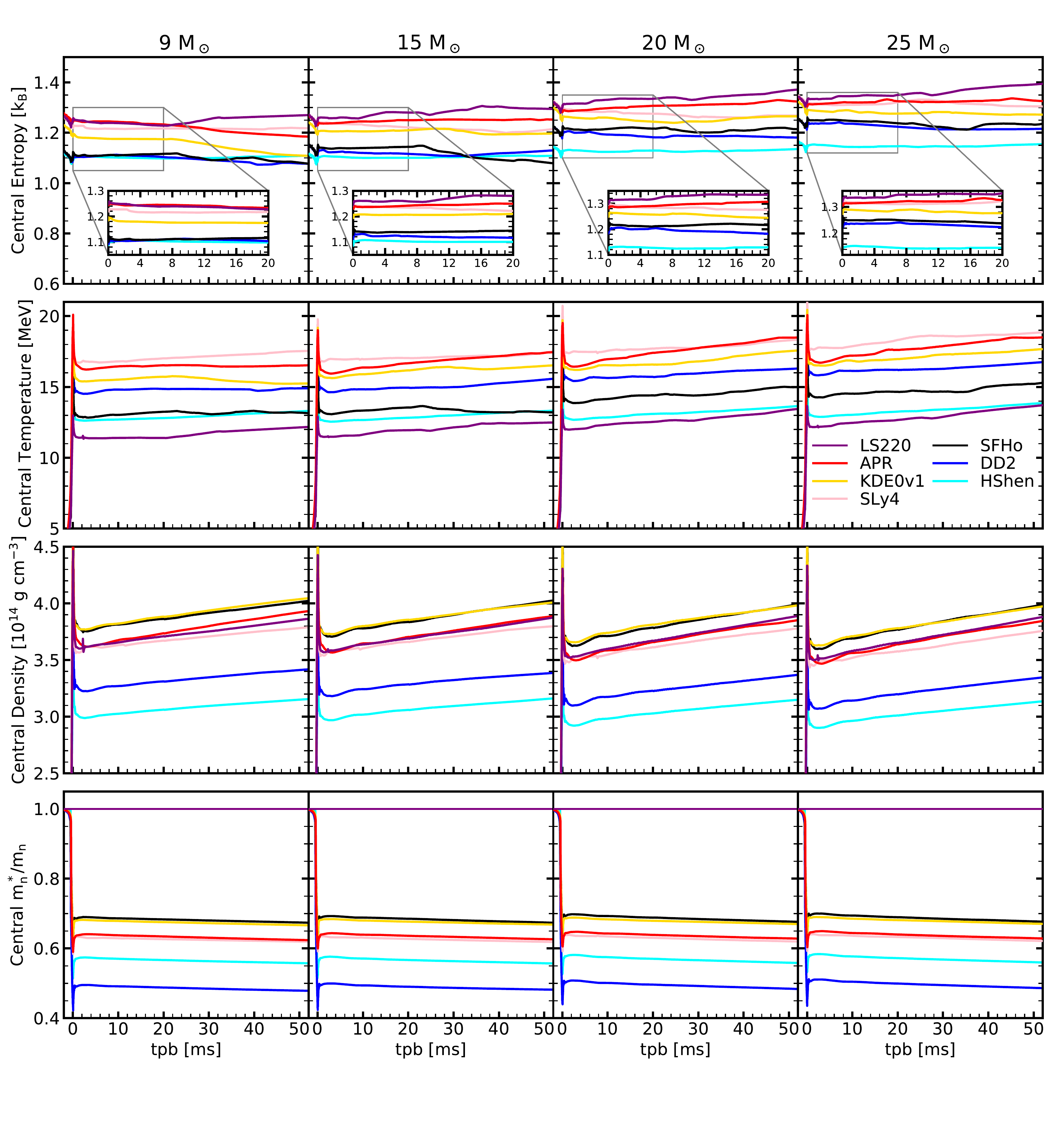}
    \caption{The first three rows show the main central thermodynamic variables for different progenitors (columns) and EOSs (colors) as a function of time post bounce. The last row shows the ratio of the nucleon effective mass to the nucleon mass calculated for the central values of densities and temperatures as a function of time post bounce. The central entropy at early times tpb $\lesssim 5$ ms is larger for EOSs that yield stronger explosions. See the text for a more detailed discussion.}
    
    \label{fig:PNS_center}
\end{figure*}

The PNS is a very high density environment, and therefore its interior can be greatly impacted by the EOS. In Figure \ref{fig:PNS_center} we show the central density, temperature and entropy per baryon as a function of time for different progenitors and EOSs. Note that the central entropy in the first 50 ms after bounce correlates extremely well with the strength of the explosion. Indeed, by comparing it to the explosion energy and shock trajectory, one can clearly see that the EOS with the highest central entropy (LS220) yields the earliest and strongest explosion, while the EOS with the lowest central entropy (HShen) does not explode at all, and after the initial stalling of the shock at $\sim 150$ km is the EOS with the quickest re-collapse. In general, the hierarchy of central entropies is the same as the hierarchy of explosion strength for all progenitors, with the possible exception of the SLy4 and KDE0v1 EOSs, which we will discuss at the end of this Section. 

As a historical note, this correlation between explodability and early-time central entropy was suggested by Hans Bethe \citep{Bethe1990_SN_review}. The explanation being the fact that entropy in the core is generated by the onset of nonequilibrium processes due to deleptonization during collapse and energy exchange between matter and the non-thermal neutrinos.  The amount of entropy generated will then affect the shock formation radius, and hence, the explodability. For example, the lower central $Y_e$ and temperature for the LS220 EoS are consistent with stronger deleptonization and larger entropy. 

Another way to illustrate the same concept, from the point of view of MLT, is that the minimum value of $\aMLT$ needed to trigger an explosion is very low (typically around 1.2/1.3, depending on the progenitor) for the EOS with the highest central entropy (LS220), while it is very large (typically around 1.7/1.8, depending on the progenitor) for the EOS with the lowest central entropy (HShen). On the other hand, central temperature and density do not correlate with the strength of explosion.

A similar relationship was found in the studies of \cite{Schneider2019} and \cite{Yasin2020_EOS_effects_1DFLASH}, who focused only on one type of EOS, i.e. a liquid drop model which uses Skyrme interactions between nucleons at high densities. With their setup they were able to vary different parameters (such as the effective mass of nucleons at saturation density, the symmetry energy, the incompressibility etc...) one by one, while keeping all the other parameters fixed. Therefore, they were able to pinpoint which property of the EOS has the largest impact on the explosion. 

Their analysis revealed that the effective mass of nuclei at saturation density $m^*_n$ is the quantity that most impacts the structure of the PNS, and therefore the explosion, since it determines how fast its contraction is \citep{Yasin2020_EOS_effects_1DFLASH} and changes the temperature of the neutrinospheres for all flavors \citep{Schneider2019}. They also pointed out that the effective mass affects the central entropy of the PNS, in that a larger effective mass would lead to a larger central entropy.

Our approach is somewhat different compared to these previous studies. We include not only EOSs based on Skyrme interactions, like the LS220, KDE0v1, and SLy4, but also the APR EOS and EOSs based upon RMF theory. The APR EOS shares the same properties as the Skyrme-type EOSs near saturation density, therefore we group them together under the label ``SRO-type". However, the nuclear potentials for the APR EOS are derived from nucleon-nucleon scattering, rather than from the energy density of nuclear matter. Alongside the EOS developed by \cite{Togashi2017_EOS_real_nuc_pot}, it is one of the few  EOSs available for supernovae simulations developed starting from realistic nuclear potentials. The EOSs calculated using RMF theory use either an NSE (SFHo and DD2) or a Thomas-Fermi (HShen) treatment of nuclei near saturation density. By considering this variety of EOS formulations  we can investigate the subtle differences among the most common frameworks used to calculate the nuclear EOS.

When we compare EOSs calculated using Skyrme interactions (i.e. LS220, KDE0v1 and SLy4), we find that the effective mass indeed correlates with the strength of explosion. However, this is not true when we compare all of the RMF-type and SRO-type EOSs, since for example the SFHo has a larger effective mass than the SLy4, but generates weaker and somewhat delayed explosions. Also, despite the APR sharing the same properties as the Skyrme-type EOSs near saturation density, it does not fit into the correlation between effective mass and strength of explosion.  For example, it has a smaller effective mass than KDE0v1 but yields stronger explosions. This is not completely unexpected since the density dependence of the effective mass is more complicated in the APR model \citep{Schneider2019_APR_EOS}. It also shows that the effective mass is not the only parameter that can predict the strength of explosion. Instead, the only quantity that seems to correlate with the strength of the explosion is the central entropy, regardless of the framework used to calculate the nuclear EOS.

A slight deviation from the correlation between explosion and central entropy is represented by the case of KDE0v1 and SLy4. These two EOSs yield shock trajectories and explosion energies that are very similar across progenitors, with KDE0v1 always giving a slightly stronger explosion. If one looks at the central entropy right after bounce, however, KDE0v1 gives a slightly smaller value than SLy4. This shows that, despite $m^*_n$ being the parameter that impacts the PNS structure the most, not surprisingly the other nuclear parameters can also change the structure of the PNS. In particular, assuming a Fermi liquid theory, one expects the central entropy to be roughly given by $S_c \sim m^*_n T_c/\rho_c^{2/3}$ \citep{Landau_Fermi_Liquid_Theory}, where $S_c$, $T_c$ and $\rho_c$ are the central entropy, temperature and density, respectively. KDE0v1 has a larger nucleon effective mass $m^*_n$, which increases the central entropy, but it also has a larger saturation density, which has the effect of increasing the central density, hence lowering the central entropy. These competing effects can slightly disrupt the hierarchy of central entropy, but since in our analysis the EOSs differ for several nuclear parameters, we don't expect a perfect correlation. Nonetheless, the correlation between central entropy right after bounce and the strength of the explosion remains quite robust.

\begin{figure*}
    \centering
    \gridline{\fig{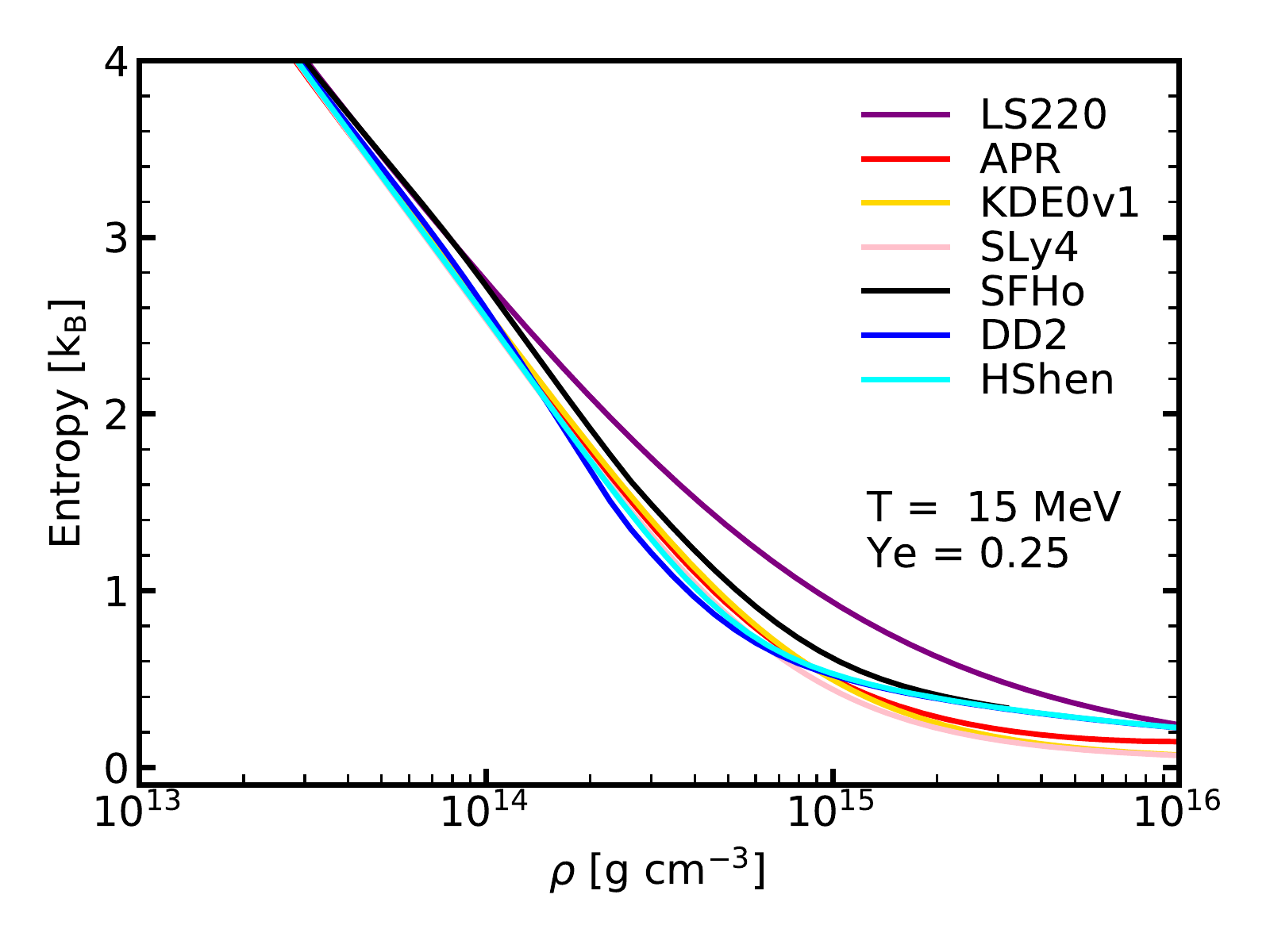}{0.5\textwidth}{(a)}
              \fig{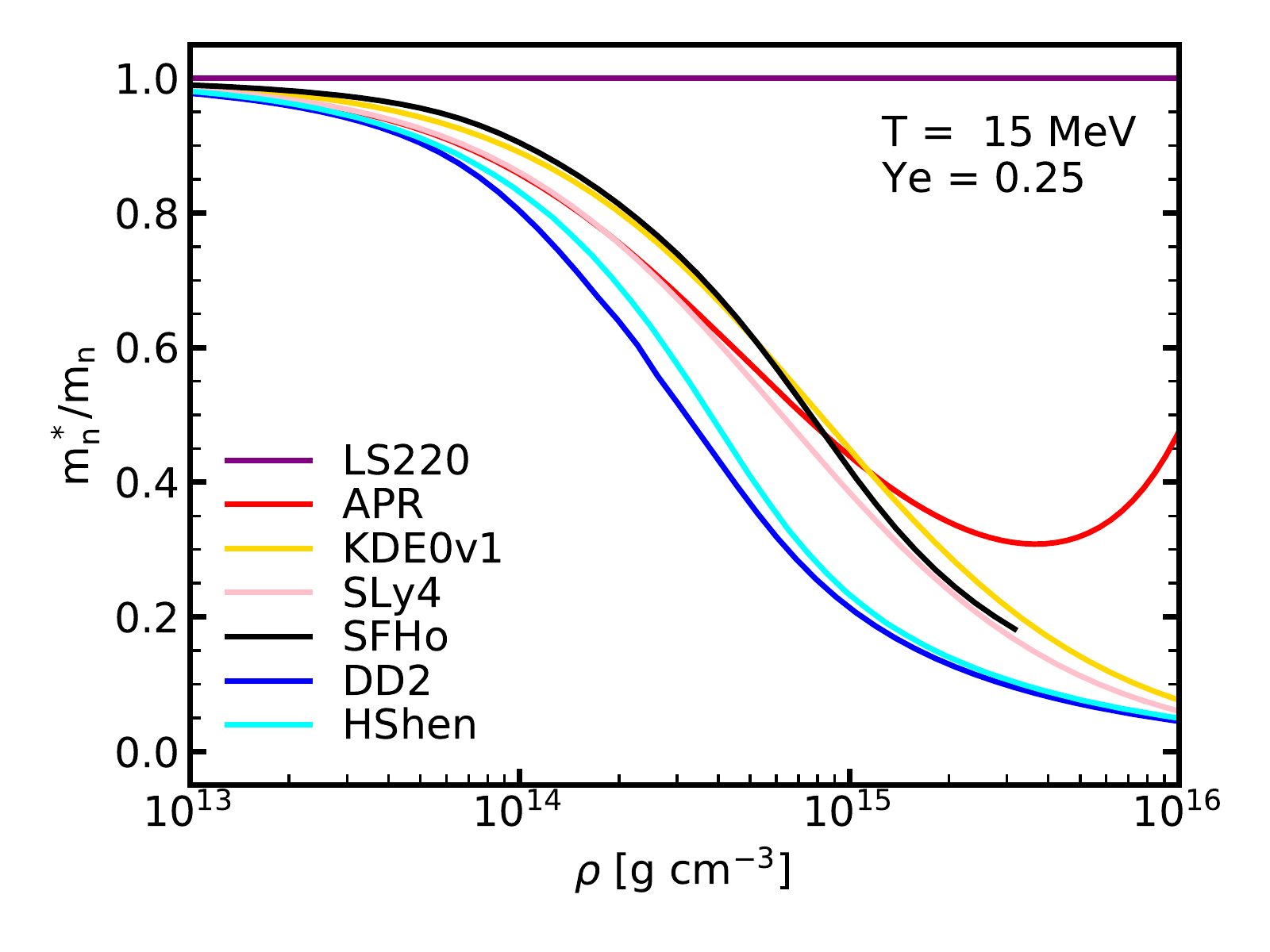}{0.5\textwidth}{(b)}}
    \caption{Panel (a) and (b) show the entropy and the effective nucleon mass, respectively, as a function of density for the EOSs considered in this paper. Temperature and electron fraction are 15 MeV and 0.25, respectively, and represent typical thermodynamic conditions reached in core-collapse supernovae environments. The nucleon effective mass is the average of the proton and neutron effective masses, which for most EOSs are different, while for HShen and LS220 are equal.}
    \label{fig:entr_meff}
\end{figure*}

It is not completely unexpected that the correlation between strength of explosion and nucleon effective mass breaks down when comparing SRO-type and RMF-type EOSs. The definition of effective mass and entropy is different in the two frameworks (and within SRO-type, it varies between the APR and the Skyrme-type EOSs).  This is especially true since relativistic effects, which are taken into account by RMF models but not by the SRO models, can be significant. As can be seen in Figure \ref{fig:entr_meff}, however, for densities between $10^{14}$ and $10^{15}$ g cm$^{-3}$ the hierarchy of effective masses is the same as the hierarchy of entropies. At first, this seems to be in contrast with the fact there is no clear correlation between effective nucleon mass at saturation density and central entropy in our simulations. However, from Figure \ref{fig:PNS_center} one can see that different EOSs yield very different thermodynamic conditions inside the PNS. Temperatures, densities and electron fractions span ranges of 10-20 MeV, $3-4\times 10^{14}$ g cm$^{-3}$, and 0.27-0.3, respectively. Therefore, comparing different EOSs at a fixed temperature and electron fraction can be misleading if one wants to analyze the impact of the EOS on the explosion of CCSNe. The bottom panel of Figure \ref{fig:PNS_center} confirms this.  Here, one can see that the hierarchy of the nucleon effective masses at the center of the newly formed PNS is different than the hierarchy of entropies, and therefore does not correlate with the strength of explosion.

\subsection{PNS interior and the role of convection}
Convection within the PNS has been shown to be very important since it can change the neutrino signal \citep{Roberts2012_PNS_cooling_convection,Mirizzi2016,Muller2020_SNReview}, the gravitational wave signal \citep{Marek2009_GW_signal_EOS,Yakunin2010_GW_SN,Muller2013_3,Andresen2017_GW_Garching,Morozova2018_GW_SN_EOS} and trigger the so-called Lepton-number Emission Self-sustained Asymmetry (LESA)  \citep{Tamborra2014}. However, its impact on the explosion is still unclear, and more high-resolution 3D simulations are needed to clarify its role. 

In a recent paper, \cite{Nagakura2020_PNS_convection} (hereafter NBRV20) analyzed the convection inside the PNS for 14 progenitors from \cite{Sukhbold2016_explodability}, with masses ranging from 9 to 60 M$_\odot$, using the SFHo EOS. Their conclusion was that there doesn't seem to be any direct influence of the shock revival on the PNS convection at early times (and vice-versa). The most significant correlation they found is between more massive PNSs and more vigorous convection. 

\begin{figure*}
    \centering
    \includegraphics[width=\textwidth]{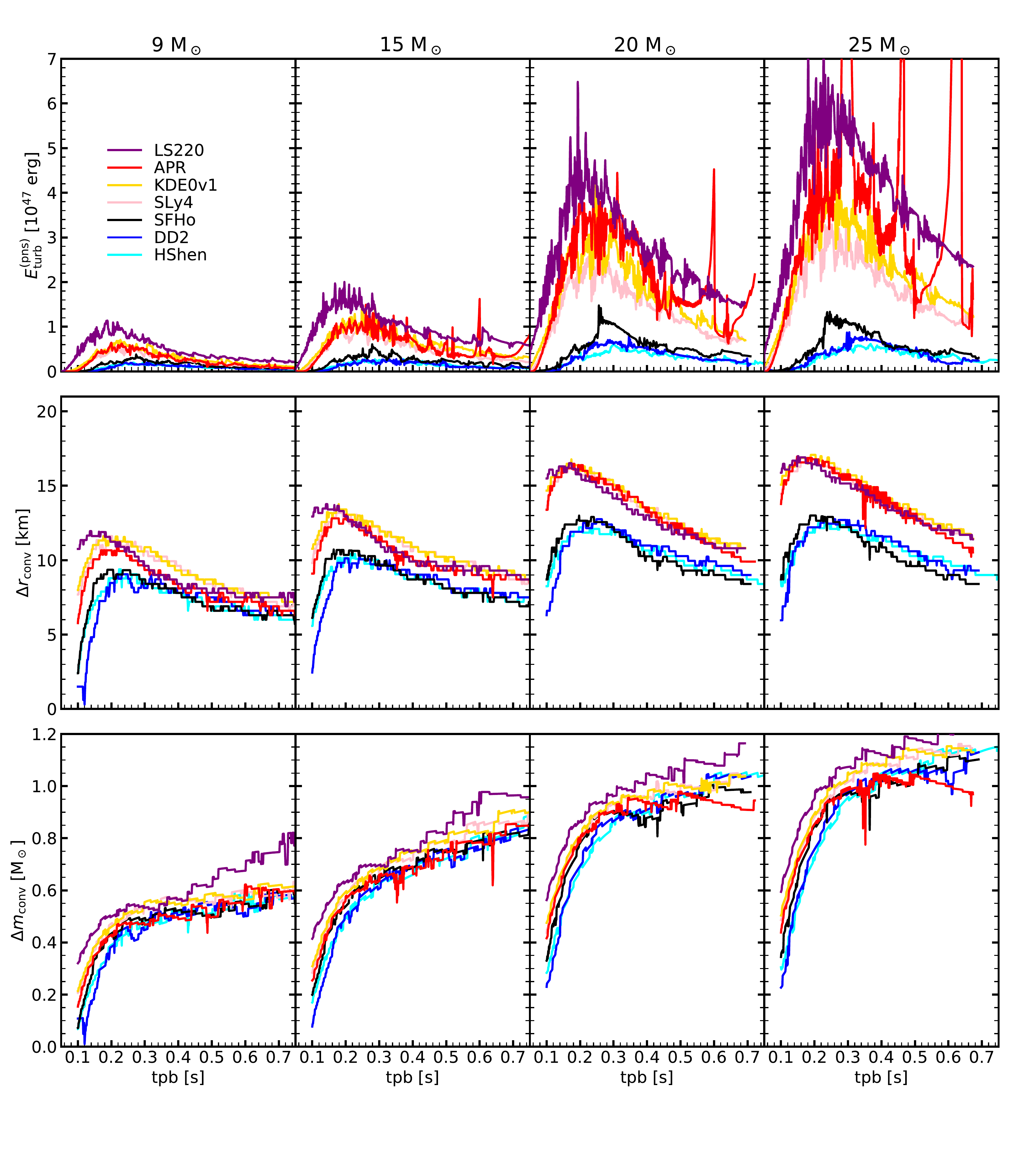}
    
    \caption{The upper panels show the total turbulent energy in the PNS as a function of time post bounce, as defined in Eq. \ref{eq:Eturb_pns}; the central panels show the width of the convective layer inside the PNS as a function of time post bounce; The bottom panels show the baryonic mass contained in the convective layer. Different columns represent different progenitors, while different colors represent different EOSs.}
\label{fig:PNS_convection_layer}
\end{figure*}

The quantity chosen to represent the ``vigor" of convection is the total turbulent energy in the PNS, defined as:

\begin{equation}
    E^{\rm pns}_{\rm turb} = \int_0^{R_{\rm pns}} \rho \vturb^2 dV ~.
    \label{eq:Eturb_pns}
\end{equation}
The top panels in Figure \ref{fig:PNS_convection_layer} show the evolution of $E^{\rm pns}_{\rm turb}$ for the various progenitors and equations of state.  

\begin{figure}
    \centering
    \includegraphics[width=\columnwidth]{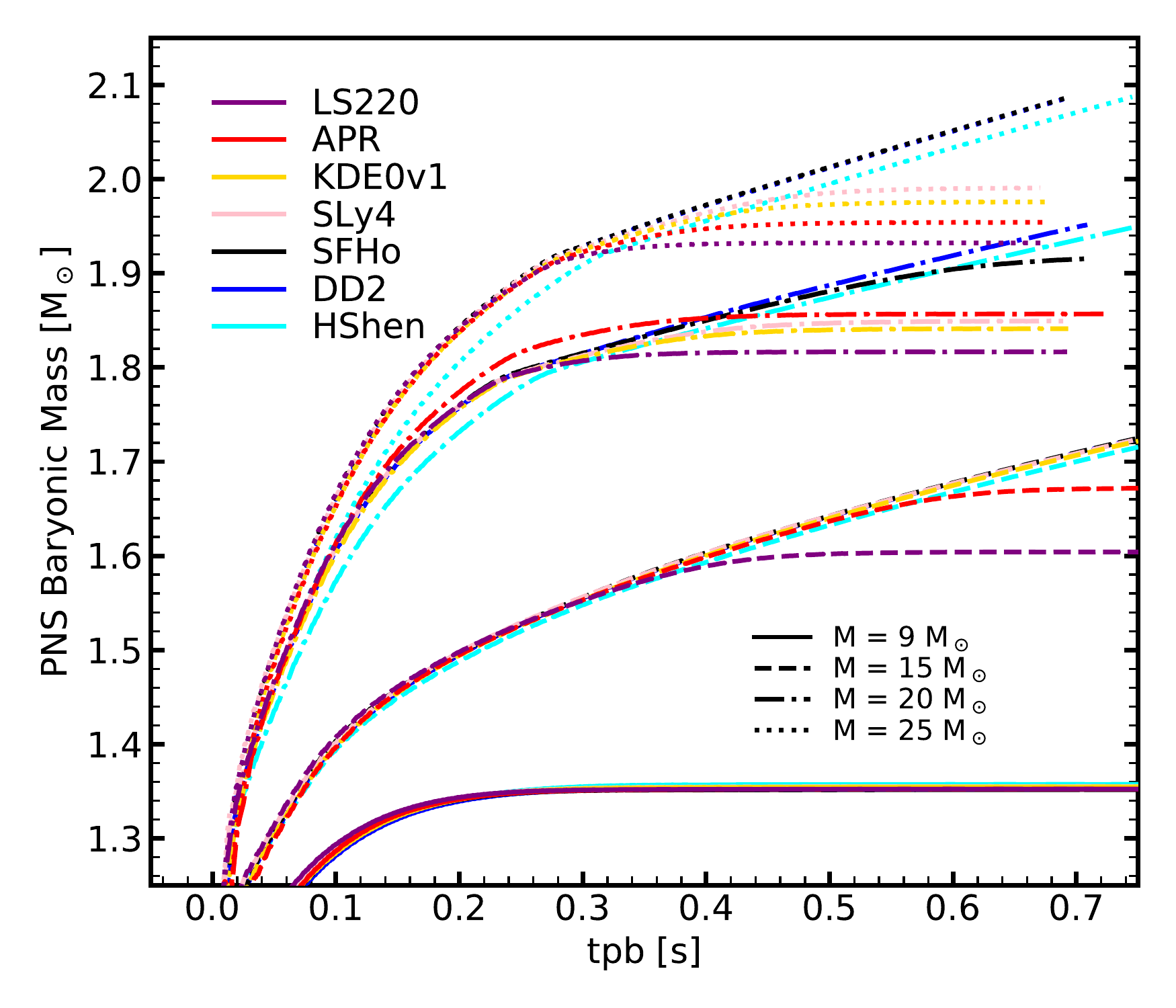}
    \caption{The baryonic mass of the PNS as a function of time post bounce. Here, different colors refer to different EOSs, while different line styles refer to different progenitors. The apparent correlation between PNS mass and progenitor mass is only a coincidence.  If more progenitors were included one would not see this correlation. The real correlation is the one discussed in the text between the PNS mass and turbulent energy.}
    \label{fig:PNS_baryonic_mass}
\end{figure}

In Figure \ref{fig:PNS_baryonic_mass} we show the time evolution of baryonic mass of the PNS for the various equations of state and progenitors considered in this paper. This figure, together with Figure \ref{fig:PNS_convection_layer}, confirms the trend found by NBRV20, i.e. that more massive PNSs show more vigorous convection. This is true for all EOSs, hence we confirm that the trend noted in the simulations of NBRV20 holds true regardless of the EOS used. 

It is also interesting to analyze each progenitor separately, and study how $E^{\rm pns}_{\rm turb}$ correlates with the strength of the explosion. For each progenitor, as can be seen in Figure \ref{fig:PNS_convection_layer}, the hierarchy of $E^{\rm pns}_{\rm turb}$ is the same as the hierarchy of the explosion energy, i.e. earlier explosions with larger explosion energies have a more vigorous PNS convection. From the analysis carried out in Section \ref{sec:PNS_interior}, one could then conclude that a larger central entropy leads to stronger convection, since both are correlated with a stronger explosion.

Another detail that emerges from our analysis is that the SRO-type EOSs seem to generate overall stronger PNS convection and form larger convective layers. By looking at the top two panels of Figure \ref{fig:PNS_convection_layer}, it is evident how there is a clear separation between the curves which represent SRO-type EOSs and the curves that represent RMF-type EOSs. 

We postpone a more thorough analysis of PNS convection and this effect in particular to future work.
The reason for this is that the turbulent convection model that we used is designed to reproduce convection in the gain region, and therefore it is not  accurate inside the PNS. In fact, by looking at the turbulent velocity near the PNS in Figure \ref{fig:calibration_vturb}, it is clear that there is a discrepancy with the 3D results. Overall, the turbulent velocities we find are roughly a factor of 10-20 smaller than in 3D.  Hence,  the turbulent energy in the PNS $E^{\rm pns}_{\rm turb}$ is also about two orders of magnitude smaller than the expectation from 3D simulations. 

As pointed out by \cite{Couch2020_STIR}, the convection inside the PNS is extremely efficient in STIR, and therefore it quickly erases the unstable thermodynamic gradients. This shuts off further growth of the turbulent velocity, which explains why in STIR it is so small. This is confirmed by the fact that, counterintuitively, smaller values of $\aMLT$ exhibit larger turbulent velocities inside the PNS.  This indicates that, for large values of $\aMLT$, convection is too efficient to allow the turbulent velocity to build-up. 

Another reason why STIR is inadequate to predict the PNS convection is that it does not take lepton gradients into account. The neutrinos trapped inside the PNS alter the full lepton gradient.  This is not merely equal to the electron fraction gradient as it is in the gain region, and as it is assumed in our STIR model. 

Calculating the full lepton gradient, however,  is not trivial \citep{Roberts2012_PNS_cooling_convection} since it involves complicated thermodynamic derivatives.  Hence, we leave a more detailed treatment of PNS convection to a future work.

Despite the inadequacy of STIR in predicting the magnitude of the turbulent velocity in the PNS, we find that it correctly predicts the width and the mass of the PNS convective layer, when compared to the results of NBRV20, and we find the same correlation between the mass of the PNS and the strength of convection.

\subsection{EOS dependence of the neutrino signal}
\label{sec:neutrinos}

\begin{figure*}[ht]
    \centering
    \includegraphics[width=0.98\textwidth]{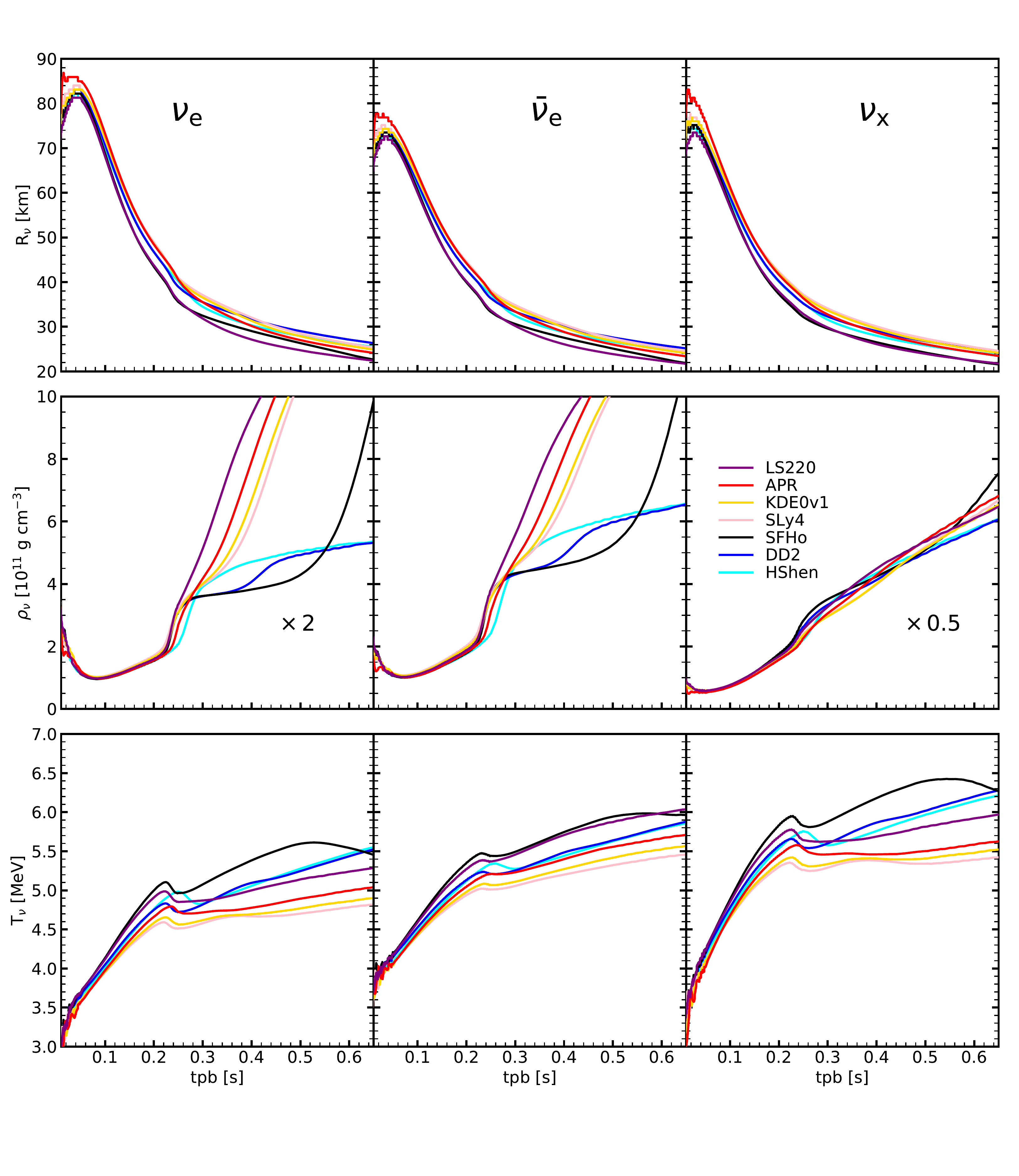}
    \caption{The top, middle and bottom panels show the radius, density and temperature of the neutrinosphere as a function of time post bounce for the 20 M$_\odot$ progenitor and the EOSs listed in Table \ref{tab:EOS_properties}, with $\aMLT = 1.48$. The curves have been smoothed out, since the sharp thermodynamic gradients lead to large jumps in density and temperature between adjacent zones, hence resulting in a saw-tooth pattern. Different columns refer to different neutrino flavors, i.e. electron neutrino $\nu_e$, electron antineutrino $\bar{\nu}_e$ and heavy lepton neutrino $\nu_x$. The densities of the electron and heavy-lepton neutrinos have been multiplied by 2 and 0.5, respectively.}
    \label{fig:nuspheres}
\end{figure*}

\begin{figure*}[ht]
    \centering
    \includegraphics[width=\textwidth]{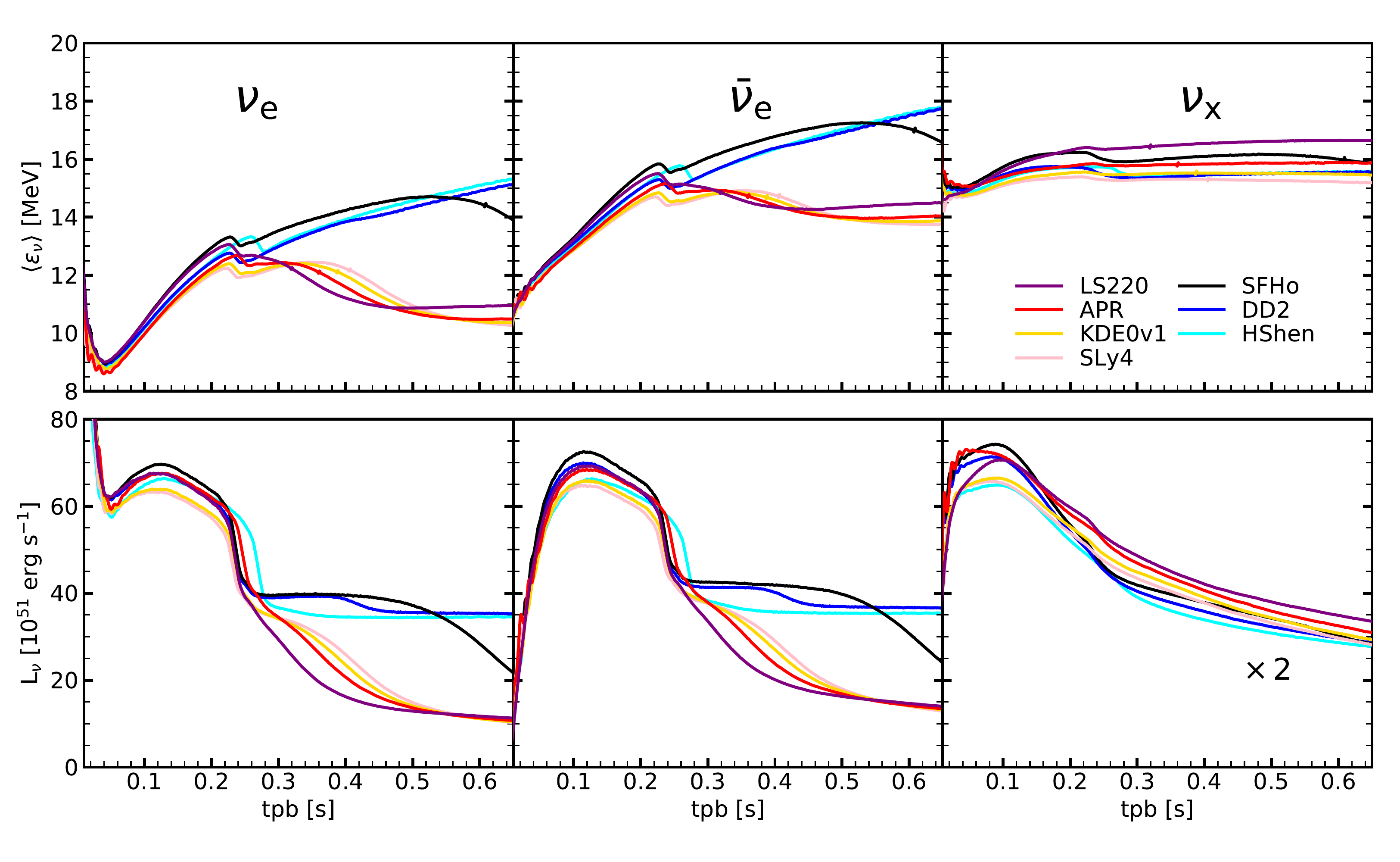}
    \caption{The top and bottom panels show the average energy and the luminosity of neutrinos, respectively, for the 20 M$_\odot$ progenitor, with $\aMLT = 1.48$, and the EOSs listed in Table \ref{tab:EOS_properties}. Both quantities are taken at 500 km from the center of the star, where neutrinos no longer interact with matter, i.e. are free streaming. The luminosity of the heavy-lepton neutrinos $\nu_X$ has been multiplied by a factor of 2 to present them on the same scale as the other neutrinos.}
    \label{fig:avee_lum_nu}
\end{figure*}

Since the pioneering work of \cite{Bethe_Wilson1985}, it has been clear that delayed neutrino heating is a primary mechanism enabling the explosion. The EOS changes the thermodynamic conditions of the PNS, and therefore it indirectly changes the neutrino-matter interaction. This in turn changes the neutrino properties, such as the luminosity evolution and the energy spectrum. The bulk of emission of neutrinos happens inside the PNS, where neutrinos decouple from matter. This is usually identified as the region in the vicinity of the neutrinosphere, defined as the layer inside the PNS at which the neutrino opacity is equal to $2/3$. Therefore, changes in temperature and density inside the PNS will affect the temperature and density profiles where the neutrinos decouple, thereby modifying the emission properties of neutrinos.

As discussed in  \cite{Schneider2019_APR_EOS}, a larger value of the effective mass at saturation changes the properties of neutrinos, since it causes a faster contraction of the PNS. This shifts the location of the different neutrinospheres to smaller radii, larger temperatures and lower densities. The only exception are the heavy-lepton neutrinos.  In this case the density of the neutrinosphere increases as the effective mass increases. It should also be noted that neutrinos of different energy decouple at different radii, since the opacity depends upon the energy-dependent neutrino interaction cross section, but for simplicity one usually refers to a single neutrino-averaged neutrinosphere, defined using an energy-averaged opacity. 

To analyze the impact of the EOS on the neutrino properties, we focus on the 20 M$_\odot$ progenitor, although a similar behavior can be seen in the other progenitors as well. The location and thermodynamic properties of the neutrinosphere for different flavors are shown in Figure \ref{fig:nuspheres}, and the luminosities and average energies are shown in Figure \ref{fig:avee_lum_nu}. 

If we limit the analysis to the SRO-type EOSs, the trends found by \cite{Schneider2019_APR_EOS} are confirmed in our simulations.  That is, the SRO-type EOSs with a larger effective mass yield a hotter neutrinosphere located at smaller radii. Similarly, larger effective masses lead to larger neutrino luminosities and energies for all flavors. If we include the RMF-type EOSs in this analysis, however, these trends are disrupted, and for example the SFHo, with a smaller effective mass at saturation than the LS220, has larger neutrinosphere temperatures, neutrino energies and luminosities for all flavors. 

In their paper, \cite{Schneider2019_APR_EOS} show the neutrino properties for non-exploding models.  In the present work, however, some of the EOSs  lead to explosions (due to the turbulent convection), and therefore  some neutrino properties will be different. In particular, as the explosion sets in, the accretion onto the PNS stops, and the neutrino emission leads to fast cooling of the inner regions.  This accelerates the contraction of the PNS and moves the neutrinospheres to regions of higher density.  In addition to this,  the temperature stops increasing and becomes roughly constant. As a consequence, the average neutrino energy also tends to a constant value, while the luminosity sharply decreases, since without accretion the neutrino production rapidly diminishes. Hence, in these simulations, the EOS is responsible for differences in the neutrino signal only in the first 150-200 ms, when turbulent convection hasn't yet triggered an explosion. After that, the different explosion dynamics will dramatically change the neutrino signal.

\begin{figure}[ht]
    \centering
    \includegraphics[width=\columnwidth]{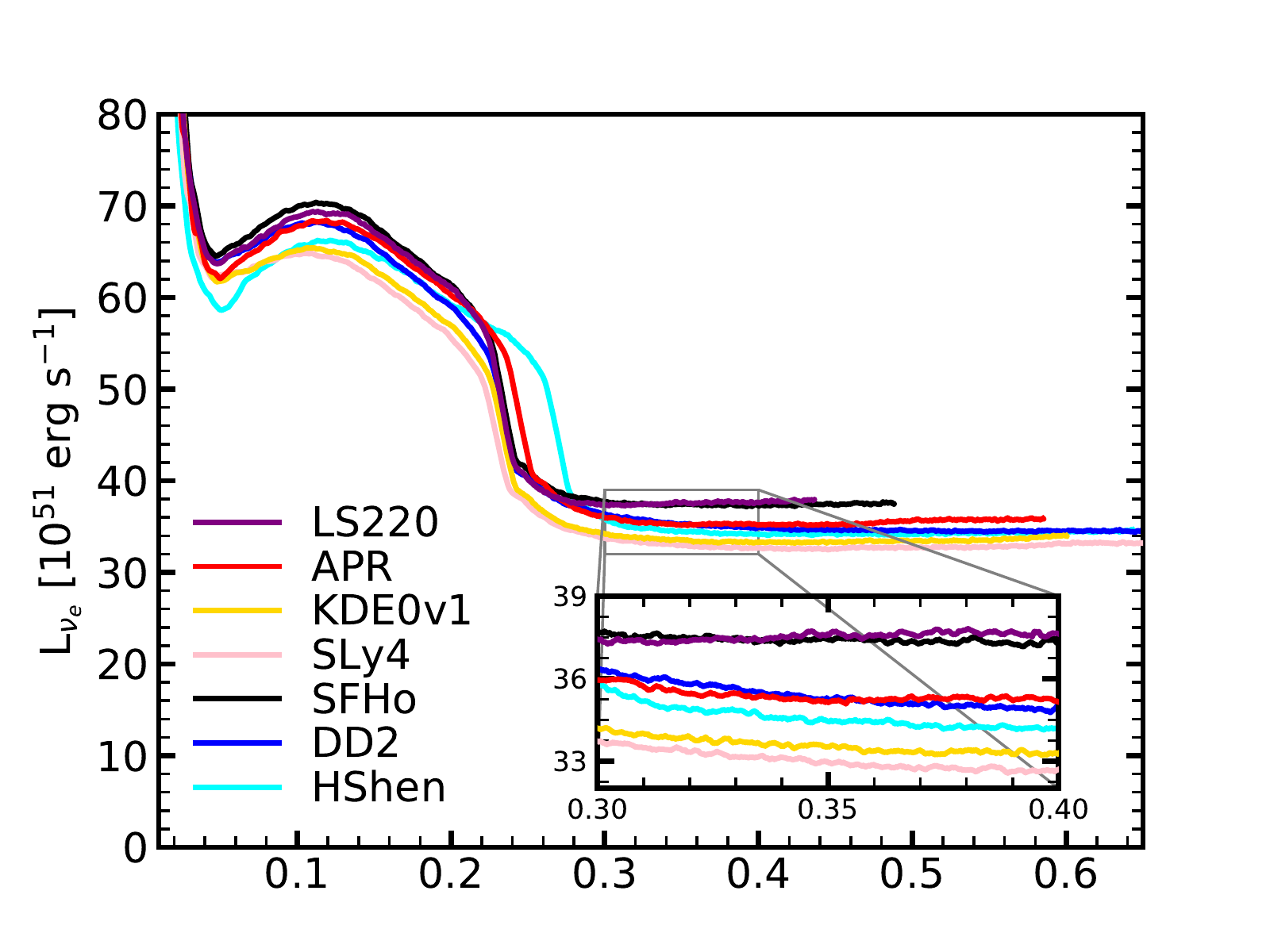}
    \caption{Electron neutrino luminosity for the 20 M$_\odot$ progenitor and the EOSs listed in Table \ref{tab:EOS_properties}, with $\aMLT = 0$ (i.e. no turbulent convection present). Highlighted is the accretion phase when the luminosity becomes constant. The hierarchy of luminosities is the same for electron neutrinos and antineutrinos, while the EOS effects on the heavy lepton neutrinos are smaller.}
    \label{fig:lum_nue}
\end{figure}

As pointed out in previous studies \citep{Hempel2012_SN_simulations,Steiner2013_SFHo} the EOS can impact the neutrino signal in several ways. For example, the presence of light nuclei near the neutrinospheres could in principle modify the neutrino emission and absorption rates, and therefore affect the neutrino signal. However, since we use the standard \cite{Bruenn1985} neutrino opacities, which don't include interactions with light nuclei, we do not see this effect in our simulations. The main quantities that can affect the neutrino luminosities and average energy are mass accretion and the location of the neutrinosphere. Larger mass accretions result in larger luminosities, whereas neutrinospheres located at smaller radii (and therefore higher temperatures) yield larger neutrino energies. A more detailed analysis of these effects can be found in \cite{Hempel2012_SN_simulations}.

It is also worth mentioning that, given the different treatment of nuclei near saturation density (namely, SNA or NSE), the fraction of heavy nuclei in inhomogeneous phases will be different in each EOS. This can also alter the neutrino signal, since a different treatment of heavy nuclei at intermediate densities will change the electron fraction (and therefore the deleptonization) of the core. Smaller electron fractions lead to more compact configurations, and therefore larger luminosities and energies (for a more detailed description, see \cite{Hempel2012_SN_simulations}). The contraction of the PNS can also be influenced by the presence of heavy nuclei below saturation density, as well as by the inclusion of general relativistic effects (taken into account only by RMF-type EOSs). Fast contractions will quickly increase the temperatures of the neutrinosphere, and therefore increase the average energies of neutrinos. A more detailed discussion can be found in \cite{Hempel2012_SN_simulations} and \cite{Steiner2013_SFHo}.

Overall, our simulations don't show a clear correlation between neutrino properties and strength of explosion. However, if one analyzes the neutrino luminosity during the accretion phase in models with no turbulence (i.e. simple spherically symmetric simulations without the inclusion of STIR) one can see the correlation between effective mass and luminosity for SRO-type models (see Figure \ref{fig:lum_nue}). On the other hand, the RMF-type models  show comparatively larger luminosities, although the hierarchy (SFHo, DD2, HShen) is the same as the hierarchy of explosion strength. Hence, we see that within SRO-type or RMF-type frameworks the correlation between luminosity and explosion strength holds.  However, this is disrupted when analyzing EOSs across both frameworks, as already pointed out in Section \ref{sec:explodability} for the case of the turbulent velocity in the gain region.

One can conclude that, compared to SRO-type EOSs, RMF-type EOSs have larger luminosities and turbulent velocities in the gain region but generate weaker explosions. The reason behind this most likely lies in the different treatment of nuclei near saturation density. A similar difference can be found in the 9 M$_\odot$ progenitor, discussed in Appendix \ref{sec:prog9}, where a different treatment of nuclear matter near saturation density leads to larger central densities which then modify the explosion energies. Hence, not only the nuclear properties, but also the treatment of nuclear matter near saturation density can affect the explosion of core-collapse supernovae.

\section{Conclusions}
\label{sec:conclusions}
In this article we have shown how the EOS can affect the explosion of CCSNe using one-dimensional, fully general relativistic simulations that employ a parametric treatment of relativistic turbulent convection. We calibrated the main free parameter of our model, $\aMLT$, by comparing our results to 3D simulations.   We then simulated the collapse of four stellar progenitors using 7 different EOSs. 

We find a remarkable correlation between the strength of the explosion and the central entropy immediately after bounce. This is in agreement with previous results obtained by \cite{Schneider2019} and \cite{Yasin2020_EOS_effects_1DFLASH}.  In these works,  EOSs calculated using Skyrme interaction models were used to analyze the impact of different nuclear parameters --- such as the effective nucleon mass, the symmetry energy and the incompressibility --- on the explosion. They concluded that the effective nucleon mass is the parameter that most strongly correlates with the explosion, as well as correlating with central entropy. Our results using the three Skyrme EOSs considered in this work confirm this. 

In addition to Skyrme EOSs, however, we have added the APR EOS and three RMF-type EOSs to the comparison. This alters the correlation between strength of explosion and effective nucleon mass. If one considers entropy and effective nucleon mass for densities between $10^{13}$ and $10^{15}$ g cm$^{-3}$ and for a fixed temperature and electron fraction, larger nucleon effective masses are correlated with larger entropies. However, in CCSNe simulations, different EOSs yield different central temperatures and densities, and therefore the correlation between effective nucleon mass and entropy breaks down. Similarly, the correlation between effective nucleon mass and strength of explosion reported in other studies \citep{Schneider2019,Yasin2020_EOS_effects_1DFLASH} is no longer present in our simulations. However, we still obtain a remarkable correlation between strength of the explosion and central entropy. Hence, we conclude that the  central entropy immediately after bounce is the best indicator of explodability, and must therefore play a key role in determining the strength of the explosion.

We also analyzed how the central entropy is correlated with  convection in the PNS. We found that more vigorous convection induces larger convective zones and is correlated with a larger central entropy. Moreover, the SRO-type EOSs generate much larger convective zones and integrated turbulent energies than RMF-type EOSs.  The role of the PNS turbulent convection should be more thoroughly investigated, possibly using non-parametric 2D and 3D simulations. Our current PNS convection, however, shows some discrepancies with 3D results, mainly in the radial profiles of the convective velocity. We therefore leave an improved treatment of the PNS convection to a future work.

Finally, we analyzed the neutrino signal for various  EOSs. We did not find any clear correlation between the strength of the explosion and neutrino luminosities or energies. However, we noticed that the RMF-type EOSs tend to produce larger neutrino luminosities and turbulent velocities in the gain region compared to the SRO-type EOSs.  This is true even despite them yielding weaker explosions. Therefore, we conclude that the approach used to calculate the EOS, and more specifically the treatment of nuclear matter near saturation density, has a significant impact on the explosion of CCSNe.
<
\acknowledgments
The authors would like to thank Sean Couch and Mackenzie Warren for fruitful discussions, Andre da Silva Schneider for his help in calculating the symmetry energy of the EOSs adopted, and Erika Holmbeck for helping us plotting the NICER constraints. Work at the University of Notre Dame supported by the U.S. Department of Energy under Nuclear Theory Grant DE-FG02-95-ER40934. EOC would like to acknowledge Vetenskapsr{\aa}det (the Swedish Research Council) for supporting this work under award numbers 2018-04575 and 2020-00452. 

\appendix 
\section{Discussion on 9 M$_\odot$ progenitor}
\label{sec:prog9}
\begin{figure*}
    \centering
    \includegraphics[width=\textwidth]{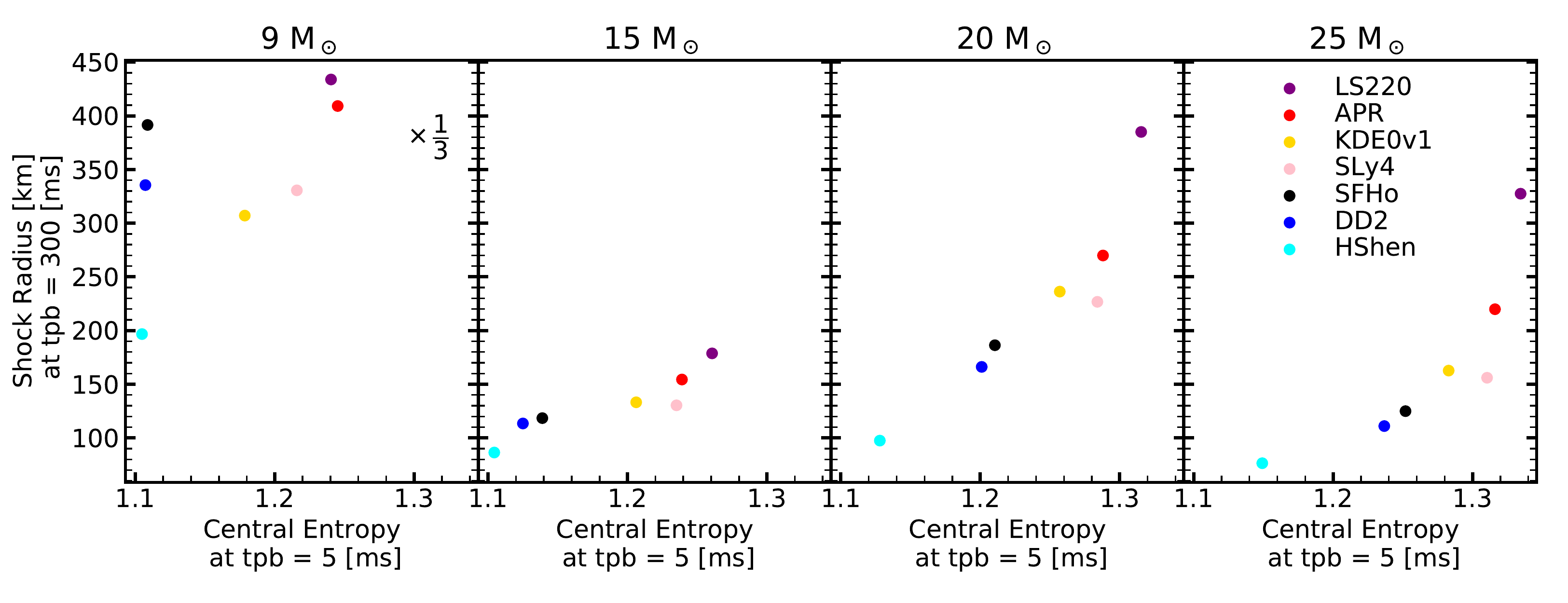}
    \caption{The vertical axis shows the radius of the shock at 300 ms post bounce.  We use this as a proxy for the strength of the explosion, since the diagnostic energy is non-zero only for exploding progenitors. The larger the shock radius, the closer the star is to an explosion. On the horizontal axis we show the value of the central entropy at 5 ms after bounce, and show that these two quantities are very well correlated (see the  text for a more detailed discussion). Different panels represent different progenitors, while different colors represent different EOSs.  Note that the shock radius for the 9 M$_\odot$ progenitor is multiplied by a factor of (1/3) to fit on the plot.}
    \label{fig:scatter_correlation}
\end{figure*}

A separate discussion for the 9 M$_\odot$ progenitor is warranted  since it does not follow the same correlations discussed in the text regarding explosion energy, central entropy and strength of the PNS convection. In particular, the DD2 and SFHo EOSs yield smaller central entropies and larger shock radii than what would be expected from the trends found in the other progenitors (see Figure \ref{fig:scatter_correlation}). To understand the origin of this discrepancy, it's useful to define the quantity.

\begin{equation}
    \Delta \rho_{\rm bounce} = \frac{\rho_{\rm bounce}^{9 M_\odot} - \rho_{\rm bounce}^{20 M_\odot}}{\rho_{\rm bounce}^{20 M_\odot}} ~.
    \label{eq:delta_rho_bounce}
\end{equation}
This represents how much larger (or smaller)  the density is at bounce than that in the  heavier, e.g. 20 M$_\odot$, progenitor. We arbitrarily chose to compare to the 20 M$_\odot$ progenitor, although the same conclusions apply if one compares to either of the other  progenitors studied in this work. 

\begin{figure*}
    \centering
    \includegraphics[width=0.5\textwidth]{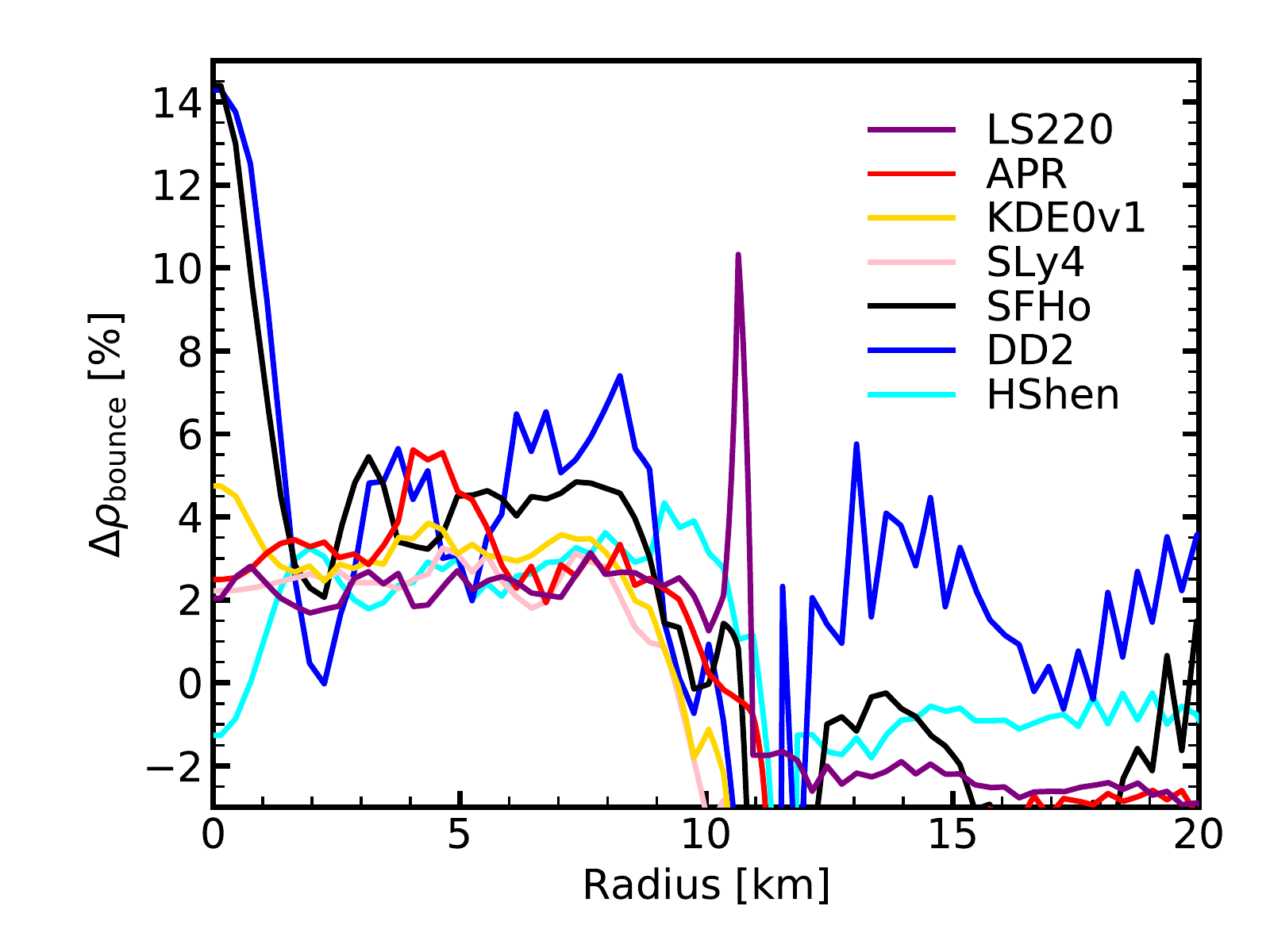}
    \caption{Relative difference between the density at bounce for the 9 M$_\odot$ and the 20 M$_\odot$ progenitors, as defined in Eq. (\ref{eq:delta_rho_bounce}), as a function of radius. In the 9 M$_\odot$ progenitor, the DD2 and SFHo EOSs yield a much larger central density than the other EOSs, for example compared to the 20 M$_\odot$ progenitor.  This is due to the different treatment of nuclear matter near the saturation density, i.e. using an NSE approach versus the SNA model.}
    \label{fig:delta_rho}
\end{figure*}

As can be seen in Figure \ref{fig:delta_rho}, the increase of the central density at bounce in the 9 M$_\odot$ progenitor with the DD2 and SFHo EOSs is much larger than with the other EOSs, which show an increase (or decrease, for the HShen) of up to  a few percent. Both the DD2 and SFHo EOSs were calculated using an NSE-approach near saturation density.  Since low-mass stars are characterized by a different core composition (mostly $^{54}$Fe, $^{40}$Ca and $^{44}$Ti, versus heavier iron-peak elements in higher-mass stars) the NSE approach changes the thermodynamic properties significantly compared to  the other EOS models.

Overall, the explosion energy for the 9 M$_\odot$ progenitor is much smaller than that of the higher mass progenitors.  This is  because the density drops off much more rapidly for low-mass stars. Hence, when the material in the outer layers is ejected, its density is very small.  As a consequence the internal and kinetic energies are decreased, making the total explosion energy smaller than that of the  higher mass stars.


\bibliographystyle{aasjournal}
\bibliography{References_SN,References_CNO,References_Nucleosynthesis,References_EOS_nuRates,References_Books}

\end{document}